\documentclass[a4paper,11pt]{article}
\pdfoutput=1 

\usepackage{jheppub} 

\usepackage{graphicx}
\usepackage{graphics}

\usepackage{amsmath}
\usepackage{booktabs}

\usepackage{slashed,xcolor}
\usepackage{multirow,array,rotating}
\usepackage{fancybox,hyperref,epstopdf}
\usepackage{morefloats}
\usepackage{color}

\def\End{\end{document}}

\def\ba{\begin{array}}
\def\ea{\end{array}}
\def\beqn{\begin{eqnarray}}
\def\eeqn{\end{eqnarray}}
\def\beqs{\begin{subequations}}
\def\eeqs{\end{subequations}}

\newcommand{\Dfb}{\mathord{\buildrel{\lower3pt\hbox{$\scriptscriptstyle{\leftrightarrow \tiny{ \ \ \ } }$}}\over {D^{\mu}}}} 

\newcommand{\Dfbd}{\mathord{\buildrel{\lower3pt\hbox{$\scriptscriptstyle\leftrightarrow$}}\over {D}_{\mu}}}

\newcommand{\lsim}{\buildrel < \over {_\sim}}
\newcommand{\gsim}{\buildrel > \over {_\sim}}

\newcommand{\MET}{\mbox{$E\kern-0.50em\raise0.10ex\hbox{/}_{T}$}}

\newcommand{\ie}{{i.e.}}

\newcommand{\bea}{\begin{eqnarray}}
\newcommand{\eea}{\end{eqnarray}}

\newcommand{\ds}{\displaystyle×}

\newcommand{\beq}{\begin{equation}}
\newcommand{\eeq}{\end{equation}}

\title{Resonant Di-Higgs Production in the $b{\bar b}WW$ Channel: Probing the Electroweak Phase Transition at the LHC}

\author[a]{T. Huang,}
\author[b,c]{J. M. No,}
\author[a]{L. Perni\'{e},}
\author[d,e]{M. Ramsey-Musolf,}
\author[a]{A. Safonov,}
\author[f]{M. Spannowsky,}
\author[d]{P. Winslow}

\affiliation[a]{Department of Physics and Astronomy, Texas A\&M University, College Station, TX 77843, USA}
\affiliation[b]{Department of Physics, King's College London, Strand, WC2R 2LS London, UK}
\affiliation[c]{Department of Physics and Astronomy, University of Sussex, Brighton BN1 9QH, UK}
\affiliation[d]{Physics Department, University of Massachusetts Amherst, Amherst, MA 01003, USA}
\affiliation[e]{Kellogg Radiation Laboratory, California Institute of Technology, Pasadena, CA 91125, USA}
\affiliation[f]{Institute of Particle Physics Phenomenology, Physics Department, Durham University, Durham DH1 3LE, UK}

\emailAdd{taohuang@physics.tamu.edu}
\emailAdd{jose$\_$miguel.no@kcl.ac.uk}
\emailAdd{lpernie@physics.tamu.edu}
\emailAdd{mjrm@physics.umass.edu}
\emailAdd{safonov@tamu.edu}
\emailAdd{michael.spannowsky@durham.ac.uk}
\emailAdd{pwinslow@physics.umass.edu}

\abstract{We analyze the prospects for resonant di-Higgs production searches at the LHC in the 
$b\bar{b} W^+ W^-$ ($W^{+} \to \ell^{+} \nu_{\ell}$, $W^{-} \to \ell^{-} \bar{\nu}_{\ell}$) channel, 
as a probe of the nature of the electroweak phase transition in Higgs portal extensions of the Standard Model.
In order to maximize the sensitivity in this final state, we develop a new algorithm 
for the reconstruction of the $b \bar{b} W^+ W^-$ invariant mass in the presence of neutrinos from the $W$ decays,
building from a technique developed for the reconstruction of resonances decaying to $\tau^{+}\tau^{-}$ pairs.
We show that resonant di-Higgs production in the $b\bar{b} W^+ W^-$ channel 
could be a competitive probe of the electroweak phase transition already with the datasets to be collected by the CMS and ATLAS 
experiments in Run-2 of the LHC. The increase in sensitivity with larger amounts of data accumulated during the High Luminosity LHC phase
can be sufficient to enable a potential discovery of the resonant di-Higgs production in this channel. 
}

\begin{document} 

\titlepage

\maketitle


\section{Motivation}
\label{sec:motivation}
With the discovery of the 
Higgs boson at the Large Hadron Collider (LHC)~\cite{Aad:2012tfa,Chatrchyan:2012xdj}, exploring the thermal history associated with electroweak 
symmetry-breaking (EWSB) has taken on heightened interest. In the Standard Model (SM), EWSB in the early Universe occurs through a 
crossover transition. In contrast,  beyond the Standard Model (BSM) scenarios may lead to a {\em bona fide}  electroweak phase transition (EWPT). 
If such a transition occurred and was both first order and sufficiently strong, it could have provided the conditions needed for generating the observed 
cosmic matter-antimatter asymmetry.

Electroweak baryogenesis (EWBG) (for recent reviews, see~\cite{Morrissey:2012db,Konstandin:2013caa}) is one of the most widely-studied and experimentally testable
scenarios for explaining the origin of the cosmic matter-antimatter asymmetry, characterized by the baryon-to-entropy density ratio $Y_B = n_B/s$ as (most precisely) 
measured by {\sc Planck}~\cite{Ade:2013zuv}
\begin{eqnarray}
Y_B = (8.59 \pm 0.11) \times 10^{-11}\, .
\end{eqnarray}
Successful baryogenesis requires three ingredients in the particle physics of the early Universe, the so-called 
``Sakharov criteria"~\cite{Sakharov:1967dj}: {\it(i)} baryon number (B)-violation; {\it(ii)} C- and CP-violation; 
{\it(iii)} departure from thermal equilibrium or a breakdown of CPT invariance. The SM contains the requisite 
B-violation in the guise of electroweak sphalerons, but it fails with regard to the last two criteria. 
CP-violation in the SM, via the CKM mixing matrix, is too feeble. In the minimal SM, the maximum Higgs mass for 
a first order EWPT is $m_h \sim 70-80$ GeV, as confirmed by a variety of theoretical Monte Carlo 
simulations~\cite{Kajantie:1996mn,Gurtler:1997hr,Laine:1998jb,Csikor:1998eu,Aoki:1999fi}, while for the observed $m_h \sim 125$ GeV 
EWSB occurred through a cross-over phase transition in the early Universe, which would not provide 
for the necessary out-of-equilibrium conditions.

In contrast, if the observed Higgs boson resides within an extended scalar sector, the nature and properties of the EWPT 
could differ significantly from those of the SM. In that case, the Universe could have undergone a strong first order EWPT 
even for a SM-like Higgs boson of mass $m_h \sim 125$ GeV. The additional scalar degrees of freedom can alter the finite temperature 
effective potential to make such a transition possible. The simplest realization of this possibility involves the extension of the SM Higgs sector 
by a single real scalar singlet $S$, the xSM~\cite{Profumo:2007wc,Barger:2007im,Espinosa:2011ax,Curtin:2014jma,Profumo:2014opa}. 
While the xSM in and of itself is unlikely to be realized in Nature, it embodies the phase transition dynamics associated with more complete models that 
contain a gauge singlet, such as the next-to-minimal supersymmetric Standard Model (NMSSM)~\cite{Huber:2006wf,Kozaczuk:2014kva}, without introducing the complications 
associated with the other degrees of freedom present in these models. As such, it provides a framework for exploring generic features of singlet-driven 
phase transitions and the corresponding low-energy phenomenology. 
To enable a strong first order EWPT in the xSM, the coupling(s) between the new scalar $S$ and the SM Higgs doublet need 
to be sizable (though still perturbative). In general, the xSM gives rise to two neutral scalars $h_{1,2}$ with masses $m_{1,2}$ that are 
mixtures of the singlet and neutral component of the doublet. The corresponding phenomenological consequences include reduced SM-like Higgs boson signal 
strengths~\cite{Profumo:2007wc,Barger:2007im,Profumo:2014opa}, deviations of the trilinear Higgs self-coupling from its SM value~\cite{Noble:2007kk,Profumo:2014opa}
and resonant di-Higgs production\cite{Dolan:2012ac,No:2013wsa}.


In this work we focus on resonant di-Higgs production at the LHC:
$pp\to h_2\to h_1 h_1$, where $h_{1}$ ($h_2$) denotes the SM-like 
(singlet-like) neutral scalars, for $m_2 > 2 \,m_1 = 250$ GeV. Within the SM, di-Higgs production is 
non-resonant, and search strategies have been proposed in $b\bar{b} \gamma \gamma$~\cite{Baur:2002qd,Baur:2003gp,Azatov:2015oxa,Kling:2016lay}, 
$b \bar{b} W^+ W^-$~\cite{Dolan:2012rv, Papaefstathiou:2012qe}, 
$b \bar{b} \tau^+ \tau^-$~\cite{Dolan:2012rv, Barr:2013tda} and $b\bar{b} b\bar{b}$~\cite{deLima:2014dta,Wardrope:2014kya,Behr:2015oqq} final states, 
all found to be very challenging due to the smallness of the non-resonant di-Higgs cross 
section~\cite{Baglio:2012np,deFlorian:2013jea,Frederix:2014hta,Borowka:2016ehy,Borowka:2016ypz}. 
Hence, ongoing di-Higgs searches by ATLAS and CMS are looking beyond the SM paradigm, also focusing on 
resonance-enhanced production mechanisms~\cite{Aaboud:2016xco,Khachatryan:2016cfa,Khachatryan:2016sey,Aad:2015xja}. 


In this context, two key issues need to be addressed. First, it is important to assess the LHC reach into the viable parameter space 
for a strong first order EWPT. There have been initial studies in this context in the $b{\bar b} \tau^+\tau^-$ final state~\cite{No:2013wsa}, 
which found that discovery at the LHC may be possible with 100 fb$^{-1}$ of integrated luminosity for relatively light $h_2$ masses, 
but a comprehensive analysis has not yet been achieved. Second, in order to achieve maximal LHC sensitivity to the xSM, 
it is crucial to determine the degree to which different di-Higgs final states provide complementary probes of $h_2$ in different regions of the possible $m_2$ mass range.
%
Here we consider the prospects for LHC discovery/exclusion of resonant di-Higgs production in the $b \bar{b} W^+ W^-$ channel, which has been 
initially studied in~\cite{Martin-Lozano:2015dja} for low $m_2$ masses ($m_2 < 500$ GeV). We cover the entire mass range 
$250$ GeV $< m_2 < 1$ TeV, focusing on what is possible to achieve with LHC Run 2. We assess the LHC potential for probing the strong first order EWPT 
parameter space by defining a set of twelve benchmark xSM parameter choices
(corresponding to twelve $h_2$ mass windows in the range $m_2 \in [250,\,850]$ GeV),
each of them giving the maximum resonant di-Higgs production cross section
[$(\sigma_{h_2}\times\mathrm{BR}_{h_2\to h_1h_1})_\mathrm{max}$] consistent with a strong first order EWPT within its mass window.

%
%
For the $b \bar{b} W^+ W^-$ analysis, we use a Multi Variate Analysis (MVA) discriminator
in order to efficiently discriminate the signal from $t\bar{t}$ production, the most important SM background
(particularly as $m_2$ increases).
Conventional experimental techniques do not allow full reconstruction of the resonance mass,
which results in diminished discrimination against the leading background.
To improve the sensitivity of the analysis,
we have deployed a novel technique for the reconstruction of the invariant mass $m_2$ in the process 
$h_2 \to h_1 h_1 \to b \bar{b} W^+ W^-$ in the presence of neutrinos from the W decays. The proposed method 
builds on the Missing Mass Calculator (MMC) technique developed for the reconstruction of resonances decaying
to $\tau^{+}\tau^{-}$ pairs~\cite{Elagin:2010aw}.
The new technique provides an estimator for $m_2$
using a likelihood constructed over the solutions of the kinematically underconstrained system and, for brevity, is referred to as the
Heavy Mass Estimator (HME) in the remainder of the paper.

We find that considering the $\ell\nu \,\ell'\nu$ ($\ell, \ell' = e,\,\mu$) final state (and assuming an eventual combination between 
the CMS and ATLAS experiments) allows to probe into the strong first order EWPT parameter space, defined by 
 [$(\sigma_{h_2}\times\mathrm{BR}_{h_2\to h_1h_1})_\mathrm{max}$], 
up to $m_2 \sim 700$ GeV with 300 fb$^{-1}$ of integrated luminosity, making this channel a promising avenue for analysis during LHC Run 2 and 
the high luminosity phase of the LHC (HL-LHC).



Our work is organized as follows: 
Section \ref{sec:model} gives an overview of the xSM, including a summary 
of current phenomenological constraints and the choice of first order EWPT-viable benchmark points.
In Section \ref{sec:bbww} we discuss the analysis of the $b{\bar b}WW$ 
channel in detail, including signal and background event generation, object reconstruction, and the 
algorithm for probabilistic reconstruction of the event kinematics. In Section \ref{sec:sensit} we apply this analysis to 
determine the LHC Run 2 and HL-LHC reach. Finally, in Section \ref{sec:out} we offer a summary and outlook.

\section{The xSM}
\label{sec:model}

\subsection{The Model} 

We consider the most general form for the xSM scalar potential that depends on a Higgs 
doublet, $H$, and real singlet, $S$ (see {\em e.g.}~\cite{Profumo:2007wc,Barger:2007im,Espinosa:2011ax}):
\begin{eqnarray}
\label{ScalarPotential1}
& V(H,S) =\ds  -\mu^2 \left( H^\dagger H \right) + \lambda \left( H^\dagger H \right)^2 + \frac{a_1}{2} \left( H^\dagger H \right) S & \nonumber \\
& \ds + \frac{a_2}{2} \left( H^\dagger H \right) S^2 + \frac{b_2}{2} S^2 + \frac{b_3}{3} S^3 + \frac{b_4}{4} S^4 . &
\end{eqnarray}
The $a_1$ and $a_2$ parameters constitute the Higgs portal, providing the only connection between the SM and the singlet scalar $S$. 
We note that in the absence of $a_1$ and the scalar self-interaction $b_3$, the potential \eqref{ScalarPotential1} has a $\mathbb{Z}_2$ symmetry that
remains exact if the singlet field does not develop a vacuum expectation value (vev). We however retain both parameters in the current study, as they play a 
leading role in the EWPT as well as in di-Higgs production at colliders. 

Boundedness of the scalar potential from below requires positivity of the quartic coefficients along all directions in field space. 
Along the $h$ ($s$) direction, this leads to the bound $\lambda > 0$ ($b_4 > 0$), while along an arbitrary direction this implies 
$a_2 > - \sqrt{\lambda\, b_4}$. After EWSB, $H \to (v_0+h) / \sqrt{2}$ with $v_0 = 246$ GeV, and we allow for a possible vev for 
$S$, \ie~$S \to x_0 + s$, where $x_0$ is taken to be positive without any loss of generality (provided that $a_1$ and $b_3$ can take either sign). 

The minimization conditions allow for two of the parameters in \eqref{ScalarPotential1} to be expressed in terms of the vevs and other 
parameters. For convenience, we choose 
\begin{eqnarray}
& \ds \mu^2 = \lambda v_0^2 + \left( a_1 + a_2 x_0 \right) \frac{x_0}{2} & \nonumber \\
& \ds b_2 = \ds - b_3 x_0 - b_4 x_0^2 - \frac{a_1 v_0^2}{4 x_0} - \frac{a_2 v_0^2}{2} .&
\label{eq:ewsb}
\end{eqnarray}
For viable EWSB, two conditions must be satisfied: ($v_0,x_0$) has to be a stable minimum, which requires
\begin{eqnarray}
\ds b_3 x_0 + 2 b_4 x_0^2 - \frac{a_1 v_0^2}{4 x_0} - \frac{ (a_1 + 2 a_2 x_0 )^2 }{ 8 \lambda } > 0 .
\end{eqnarray}
Furthermore, the electroweak minimum must be the absolute minimum, which we impose numerically. 
After EWSB, the Higgs portal parameters $a_1, a_2$ and the singlet vev $x_0$ induce a mixing between the states $h$ and $s$. The mass-squared matrix entries are 
\begin{eqnarray}
& m_{h}^2 \equiv \ds \frac{d^2 V}{dh^2} = 2 \lambda v_0^2 & \nonumber \\
& m_{s}^2 \equiv \ds \frac{d^2 V}{ds^2}  = b_3 x_0 + 2 b_4 x_0^2 - \frac{a_1 v_0^2}{4 x_0} & \nonumber \\
& m_{hs}^2 \equiv \ds \frac{d^2 V}{dh ds} = \left(a_1 + 2 a_2 x_0 \right) \frac{v_0}{2} .
\label{mixingM}
\end{eqnarray}
with the corresponding eigenvalues given by 
\begin{eqnarray}
& m_{1,2}^2 = \ds \frac{ m_{h}^2 + m_{s}^2 \mp \left| m_{h}^2 - m_{s}^2 \right| \sqrt{ 1 + \ds \left( \frac{ m_{hs}^2 }{ m_{h}^2 - m_{s}^2 } \right)^2 } } {2} ,&
\label{Meigenvalues}
\end{eqnarray}
with $m_2 > m_1$ by construction. The mass eigenstates are given by
\begin{eqnarray}
& h_1 = h \cos \theta + s \sin \theta & \nonumber \\
& h_2 = - h \sin \theta + s \cos \theta &
\label{eigstates}
\end{eqnarray}
where we identify the more $SU(2)_L$-like state $h_1$ with the  Higgs boson observed at the LHC~\cite{Aad:2012tfa, Chatrchyan:2012xdj} by setting 
$m_1 = 125$ GeV, and where  $h_2$ is a singlet-like mass eigenstate. The mixing angle $\theta$ is defined as 
\begin{eqnarray}
\sin 2 \theta =  \ds \frac{ \left( a_1 + 2 a_2 x_0 \right) v_0  }{ \left( m_1^2 - m_2^2 \right) } .
\label{sin2theta}
\end{eqnarray} 
By virtue of~\eqref{eigstates}, the couplings of $h_1$ and $h_2$ to SM vector bosons and fermions are universally rescaled w.r.t. the SM Higgs couplings, 
\begin{eqnarray}
g_{h_1 xx} = c_{\theta} \; g_{h xx}^{\mathrm{SM}}\quad \quad , \quad \quad g_{h_2 xx} = s_{\theta} \; g_{h xx}^{\mathrm{SM}}
\end{eqnarray}
with $xx$ representing a SM final state different from $xx = hh$, and $c_{\theta}$, $s_{\theta} \equiv \cos \theta$, $\sin \theta$.
In addition to these couplings, the tri-scalar interactions will play an important role in the following discussion of di-Higgs production. Of particular interest are the  interactions
$\lambda_{211}h_2\,h_1\,h_1$ and $\lambda_{111}\,h_1\,h_1\,h_1$, which follow from \eqref{ScalarPotential1} after EWSB, with
\bea
\label{g211}
\lambda_{211} &=& \frac{1}{4}\left[ (a_1+2a_2x_0)\, c_{\theta}^3 + 4 v_0 (a_2 -3 \lambda)\, c_{\theta}^2 s_{\theta} \right. \nonumber\\
 &-& \left.2(a_1+2a_2x_0 -2b_3-6b_4x_0)\, c_{\theta} s^2_{\theta} -2a_2 v_0 \, s^3_{\theta} \right] \\
\lambda_{111} &=& \lambda v_0 \, c_{\theta}^3 + \frac{1}{4} (a_1+2a_2x_0) \, c_{\theta}^2 s_{\theta} +
\frac{1}{2} a_2v_0\, c_{\theta} s^2_{\theta} + \left( \frac{b_3}{3} + b_4 x_0 \right) s^3_{\theta} \nonumber
\eea

\subsection{Current Phenomenological Constraints}
\label{Section_Constraints_Pheno}

\begin{figure}[h!]
\centering
\includegraphics[width=0.75\textwidth]{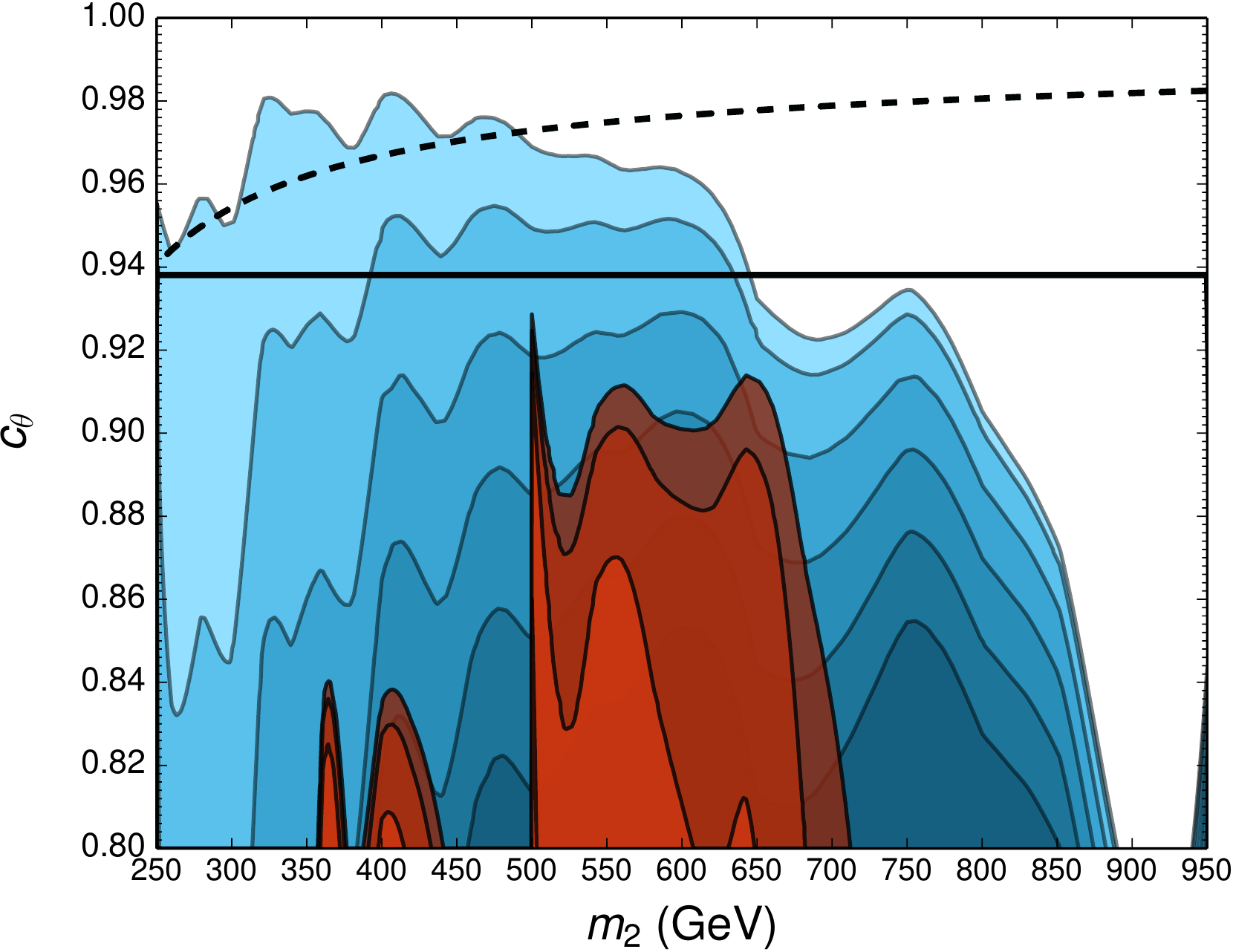}
\caption{Present 95\% C.L. excluded regions from 
Public ATLAS and CMS $p p \to h_2 \to Z Z$ (blue)~\cite{Aad:2015kna,Aad:2015ipg,Chatrchyan:2013yoa,Khachatryan:2015cwa} 
and $p p \to h_2 \to h_1 h_1$ (red)~\cite{Aaboud:2016xco,Khachatryan:2016cfa,Khachatryan:2016sey,Aad:2015xja} searches in the ($m_2,\, c_{\theta}$) plane, 
for different values of $\lambda_{211}/v_0$. The colour gradient for $p p \to h_2 \to Z Z$ corresponds to values 
of $\lambda_{211}/v_0 = 0,\,1, \,2,\,3,\,4,\,5$ (from lighter to darker) 
while the colour gradient for $p p \to h_2 \to h_1 h_1$ corresponds to values of $\lambda_{211}/v_0 = 3,\,4,\,5$ (from lighter to darker). 
Also shown are the 95\% C.L. lower limit on $c_{\theta}$ from Higgs signal strength measurements (horizontal solid-black line)~\cite{Aad:2015pla} and the 
95\% C.L. lower limit on $c_{\theta} (m_2)$ from EWPO (dashed-black line), both being independent of $\lambda_{211}/v_0$.}
\label{Fig_Limits}
\end{figure}

The singlet-doublet mixing $s_{\theta}$ is constrained by measurements of Higgs signal strengths, since all the signal rates 
associated with Higgs measurements get rescaled by $c_{\theta}^2$. Currently, the limit from LHC Run 1 data is $s_{\theta}^2 \leq 0.12$ at 95\% C.L.~\cite{Aad:2015pla}. 
In addition, ATLAS and CMS searches for a heavy SM-like Higgs boson provide a probe of $h_2$. For $m_2> 2\,m_1$, which we focus on in this work, 
the decay mode $h_2 \to h_1 h_1$ is kinematically allowed, with a partial width given by
\begin{eqnarray}
\Gamma_{h_2 \to h_1 h_1} = \frac{\lambda_{211}^2 \sqrt{1 - \ds 4\, m_1^2 / m_2^2 } }{8 \pi m_2} . 
\label{partialWidthh1h1}
\end{eqnarray}
Defining $\Gamma^{\mathrm{SM}}(m_2)$ as the SM Higgs width evaluated at $m_2$ (as given {\it e.g.} in~\cite{Heinemeyer:2013tqa}), the total 
width for the $h_2$ boson is given by 
\begin{eqnarray}
\Gamma_{h_2} = s^2_{\theta} \; \Gamma^{\mathrm{SM}}(m_2) + \Gamma_{h_2 \to h_1 h_1}
\end{eqnarray}
and the signal strength (normalized to the SM value for $m_h = m_2$) for $p p \to h_2 \to x x$ is
\begin{eqnarray}
\mu^{xx}_{h_2} = s^4_{\theta}\,\frac{\Gamma_{xx}^{\mathrm{SM}}(m_2)}{ \Gamma_{h_2} } \, .
\label{h2signalstrength}
\end{eqnarray}
By means of~\eqref{partialWidthh1h1}-\eqref{h2signalstrength}, we can then express the production cross sections $p p \to h_2 \to V V$ (with $ V = W,\,Z$ gauge bosons) 
and $p p \to h_2 \to h_1 h_1$ as 
\begin{eqnarray}
\sigma_{V V} = \sigma^{\mathrm{SM}}(m_2) \times s_{\theta}^4 \, \frac{\mathrm{BR}^{\mathrm{SM}}_{VV}(m_2)}{s_{\theta}^2 + \frac{\lambda_{211}^2}{v_0^2} 
\,f(m_2)} \,\,\, , \,\,\,  
\sigma_{h_1 h_1} = \sigma^{\mathrm{SM}}(m_2) \times s_{\theta}^2 \, \frac{ \frac{\lambda_{211}^2}{v_0^2} \,f(m_2)  }{s_{\theta}^2 + \frac{\lambda_{211}^2}{v_0^2} \,f(m_2)}
\label{VVh1h1XS}
\end{eqnarray}
with $\sigma^{\mathrm{SM}}(m_2)$ being the SM Higgs LHC production cross section and $\mathrm{BR}^{\mathrm{SM}}_{VV}(m_2)$ the SM Higgs branching fraction into $VV$
for $m_h = m_2$, and $f(m_2)$ given by
\begin{eqnarray}
f(m_2) = \frac{v_0^2\, \sqrt{1 - \ds 4\, m_1^2 / m_2^2 }}{8 \pi m_2\,\Gamma^{\mathrm{SM}}(m_2)} \, .
\end{eqnarray}
We note that heavy Higgs searches in other final states ({\it e.g.} $h_2 \to \bar{f} f$) are much 
more challenging than the ones discussed above, and are disregarded in what follows.
The ATLAS and CMS Collaborations have performed searches for heavy Higgs bosons both in the 
$h_2 \to V V$~\cite{Aad:2015kna,Aad:2015ipg,Chatrchyan:2013yoa,Khachatryan:2015cwa} and  
$h_2 \to h_1 h_1$ with $h_1 h_1 \to b \bar{b} b \bar{b}$~\cite{Aaboud:2016xco,Khachatryan:2016cfa} and $h_1 h_1 \to b \bar{b} \gamma\gamma$, 
$h_1 h_1 \to b \bar{b} \tau \tau$~\cite{Khachatryan:2016sey,Aad:2015xja}. In Fig.~\ref{Fig_Limits} we show the 95\% C.L. limits from these searches 
in the ($m_2,\,c_{\theta}$) plane for increasing values of $\lambda_{211}/v_0$, using~\eqref{VVh1h1XS}.

\vspace{1mm}

It is important to note that the limits shown in Fig.~\ref{Fig_Limits} are derived from present, publicly 
available (published) analyses, and better analysis techniques can make the $h_2 \to h_1 h_1$ search channel more 
competitive relative to the $h_2 \to Z Z$ one, as compared to present bounds. We also stress that 
a discovery in any one channel would not allow by itself to individually measure $c_{\theta}$ and $\lambda_{211}/v_0$.
This highlights the need to explore various channels and final states, like $b \bar{b} W W$, to correctly 
interpret a potential discovery of a new state $h_2$.

We now turn to the discussion of the constraints on $m_2$ and $c_{\theta}$ from electroweak precision observables (EWPO). 
The effects of the xSM on EWPO may be accurately characterized by its modification of the oblique parameters $S$, $T$, and $U$
w.r.t.~the SM. From~\eqref{eigstates}, the shift in an oblique parameter $\mathcal{O}$ can be written entirely in terms of the 
SM Higgs contribution to that parameter, $\mathcal{O}^{\mathrm{SM}}(m)$ (which can be found {\em e.g.}~in~\cite{Peskin:1991sw,Hagiwara:1994pw}), 
where $m$ is either $m_1$ or $m_2$. These shifts then take the form
\begin{eqnarray}
\label{oblique_1}
\Delta \mathcal{O} = (c^2_{\theta} -1) \mathcal{O}^{\mathrm{SM}}(m_1) + s^2_{\theta} \; \mathcal{O}^{\mathrm{SM}}(m_2) 
= s^2_{\theta} \left[ \mathcal{O}^{\mathrm{SM}}(m_2) - \mathcal{O}^{\mathrm{SM}}(m_1) \right] \, . 
\end{eqnarray}
The best-fit values for the shifts $\Delta \mathcal{O}^0_i$ and standard deviations $\sigma_i$ from the most recent post-Higgs-discovery electroweak fit to the SM by the 
Gfitter group~\cite{Baak:2014ora} (for $U = 0$, which is a very accurate approximation in the xSM) are given by
\begin{equation}
 \label{chi_EWPO1}
\begin{array}{c}
\Delta S \equiv S - S_{\mathrm{SM}} = 0.06 \pm 0.09 \\
\Delta T \equiv T - T_{\mathrm{SM}} = 0.10 \pm 0.07
\end{array} \quad \quad \quad \quad 
\rho_{ij} = \left(\begin{array}{cc}
             1 & 0.91\\
             0.91 &1
            \end{array}\right)
\end{equation}
being $\rho_{ij}$ the covariance matrix in the $S-T$ plane.
We then perform a $\Delta\chi^2$ fit to obtain the 95\% C.L. allowed region in the ($m_2,\, c_{\theta}$) plane 
\begin{equation}
 \label{chi_EWPO2}
\Delta\chi^2_{\mathrm{EW}}(m_2,c_{\theta}) = \sum_{i,j} \left[\Delta\mathcal{O}_{i}(m_2,c_{\theta}) - \Delta\mathcal{O}^0_{i}\right] (\sigma^2)_{ij}^{-1}
\left(\Delta\mathcal{O}_{j}(m_2,c_{\theta}) - \Delta\mathcal{O}^0_{j}\right)\, ,
\end{equation}
where $\Delta\mathcal{O}^0_{i}$ denote the central values in (\ref{chi_EWPO1}) and $(\sigma^2)_{ij} \equiv \sigma_i \rho_{ij} \sigma_j$, being $\sigma_i$ the $S$ and 
$T$ standard deviation from (\ref{chi_EWPO1}). 
The $95\%\, \mathrm{C.L.}$ exclusion limit from $\Delta\chi^2_{\mathrm{EW}}(m_2,c_{\theta}) = 5.99$ is shown in Fig.~\ref{Fig_Limits}.

%
%
%
%
%

\subsection{The Electroweak Phase Transition: Benchmarks for $h_2 \to h_1 h_1$ Production}

The character of the EWPT is understood in terms of the finite temperature effective potential, $V_{\rm eff}^{T\neq 0}$ 
(see~\cite{Quiros:1994dr} for a review). It is well-known that the standard derivation of $V_{\rm eff}^{T\neq 0}$ suffers from gauge 
dependence\footnote[1]{The value of the EWSB vev at the critical temperature, $\phi (T_c)$, is inherently gauge-dependent as it is not an observable. 
Furthermore the standard method for extracting $T_c$ also introduces a separate and spurious gauge-dependence. The consequence is that the conventional 
criterion for avoiding baryon washout during a first order 
EWPT (which defines a ``strong" first order EWPT), $\phi(T_c) / T_c \gtrsim 1$, inherits both sources.}~\cite{Patel:2011th}, 
and here we employ a high-temperature expansion to restore gauge-independence to our analysis (see~\cite{Profumo:2014opa} for details). Doing so
requires considering the $T=0$ Coleman-Weinberg 1-loop effective potential and retaining only the gauge-independent thermal mass corrections 
to $V_{\rm eff}^{T\neq 0}$, which are essential for high-temperature electroweak symmetry restoration. 
This limit is particularly well-suited to the xSM, which generates the barrier between the broken and unbroken electroweak phases required for a first order EWPT
at tree-level via the parameters $a_1$ and $b_3$ in~\eqref{ScalarPotential1}.

In the high-temperature limit, we follow~\cite{Pietroni:1992in,Profumo:2007wc} and write the $T$-dependent, gauge-independent (indicated by the presence of a bar) vevs in a 
cylindrical coordinate representation as
\begin{equation}
\bar{v}(T) / \sqrt{2} = \bar{\phi} \cos \alpha (T), \quad \bar{x}(T) = \bar{\phi} \sin \alpha (T)\, .
\end{equation}
with $\bar{v}(T=0) = v_0$ and $\bar{x}(T=0) = x_0$. 
The critical values $\bar{\phi}(T_c)$ and $\alpha(T_c)$ are determined by minimizing 
$V_{\rm eff}^{T\neq 0} (\phi, \alpha, T)$, while $T_c$ is defined as the temperature at which the broken and unbroken phases are degenerate:
$V_{\rm eff}^{T \neq 0} (\phi, \alpha \neq \pi/2, T_c) = V_{\rm eff}^{T \neq 0} (\phi, \alpha = \pi/2, T_c)$.
A strong first order EWPT is defined by a sufficient quenching of the sphaleron transitions in the broken 
electroweak phase (see {\it e.g.}~\cite{Morrissey:2012db} for details). The energy of the 
electroweak sphaleron is proportional to the SU(2)$_L$-breaking energy scale, $\bar{v}(T)$, and as such 
the approximate criterion for a strong first order EWPT is then $\cos \alpha (T_c ) \,\bar{\phi} (T_c)/T_c \gtrsim 1$. 

With these considerations in mind,  we implement the xSM in the high-temperature 
limit in {\sc CosmoTransitions}~\cite{Wainwright:2011kj} to obtain numerically all above quantities characterizing the EWPT and calculate 
the finite temperature thermal tunneling rate into the electroweak phase; the latter must be sufficiently fast in order to preclude the 
possibility of the Universe becoming stuck in a false metastable phase. Taking $a_1$, $b_3$, $x_0$, $b_4$ and $\lambda$ as our
independent parameters (the remaining two are fixed by the values of $v_0$ and $m_h$), we perform a MC scan of the xSM parameter space within the following ranges
\begin{equation}
a_1/\text{TeV}, b_3/\text{TeV} \in [-1, 1], \quad x_0/\text{TeV} \in [0, 1], \quad b_4, \lambda \in [0, 1]
\end{equation}
where the lower bounds on the quartic couplings $b_4$ and $\lambda$ ensure vacuum stability. With our choice of independent  parameters, 
$c_{\theta}$, $a_2$ and $m_2$ are fixed by the parameters of the scan. We impose a na\"ive perturbativity bound
on the Higgs portal coupling $a_2/2 \lesssim 5$ \cite{Curtin:2014jma}. We require compatibility with the various experimental constraints 
discussed in Section~\ref{Section_Constraints_Pheno},
and demand a strong first order EWPT as described above, together with a sufficient tunneling rate.

\begin{table*}[t!]
\resizebox{\columnwidth}{!}{%
  \begin{tabular}{| c | c | c | c | c | c | c | c | c| c| c| c | c| c| c |}
    \hline
 & $\cos \theta$ & $m_2$ & $\Gamma_{h_2}$ & $x_0$ & $\lambda$ & $a_1$ & $a_2$ & $b_3$ & $b_4$ & $\lambda_{111}$ & $\lambda_{211}$ & $\sigma$ & BR ~ \\         
           &               & (GeV) &     (GeV)        & (GeV) &           & (GeV) &       & (GeV) &       &      (GeV       &       (GeV)     &  ~  (pb) ~  &  \\         
    \hline
B1 & 0.961 & 258 &  0.68 & 307 & 0.52 & -266 & 0.26 & -138 & 0.26 & 110 & -94.6 & 1.19 &  0.50 \\
B2 & 0.976 &  341 & 2.42 &  257 & 0.92 & -377 & 0.39 & -403 & 0.77 & 204 & -150 & 0.59 & 0.74 \\
B3 & 0.982 &  353 & 2.17 &  265 & 0.99 & -400 & 0.45 & -378 & 0.69 & 226 & -144 & 0.44 & 0.76 \\
B4 & 0.983 &  415 & 1.59 &  54.6 & 0.17 & -642 & 3.80 & -214 & 0.16 & 44.9 & 82.5 & 0.36 & 0.33 \\
B5 & 0.984 &  455 & 2.08 &  47.4 & 0.18 & -707 & 4.63 & -607 & 0.85 & 46.7 & 93.5 & 0.26 & 0.31 \\
B6 & 0.986 &  511 & 2.44 &  40.7 & 0.18 & -744 & 5.17 & -618 & 0.82 & 46.6 & 91.9 & 0.15 & 0.24 \\
B7 & 0.988 &  563 & 2.92 &  40.5 & 0.19 & -844 & 5.85 & -151 & 0.08 & 47.1 & 104 & 0.087 & 0.23 \\
B8 & 0.992 &  604 & 2.82 &  36.4 & 0.18 & -898 & 7.36 & -424 & 0.28 & 45.6 & 119 & 0.045 & 0.30 \\
B9 & 0.994 &  662 & 2.97 &  32.9 & 0.17 & -976 & 8.98 & -542 & 0.53 & 44.9 & 132 & 0.023 & 0.33 \\
B10 & 0.993 &  714 & 3.27 &  29.2 & 0.18 & -941 & 8.28 & 497 & 0.38 & 44.7 & 112 & 0.017 & 0.20 \\
B11 & 0.996 &  767 & 2.83 &  24.5 & 0.17 & -920 & 9.87 & 575 & 0.41 & 42.2 & 114 & 0.0082 & 0.22 \\
B12 & 0.994 &  840 & 4.03 & 21.7 & 0.19 & -988 & 9.22 & 356 & 0.83 & 43.9 & 83.8 & 0.0068 & 0.079 \\
 \hline
  \end{tabular}
  }
  \caption{\small Values of the various xSM independent and dependent parameters for each of the benchmark values chosen to {\bf maximize} 
  the $\sigma_{h_2}\times\mathrm{BR}_{h_2\to h_1h_1}$ value at the LHC.}
\label{T1}
\end{table*}

From the results of our scan, we define twelve consecutive $h_2$ mass windows, each 50 GeV wide (starting from the $h_1 h_1$ production 
threshold $m_2 = 250$ GeV), which together span the range $m_2 \in [250,\,850]$ GeV. The upper bound $m_2 = 850$ GeV is determined by the fact that the scan does not yield 
experimentally viable points compatible with a strong first order EWPT above $m_2 \sim 850$ GeV even though it potentially accepts points up to $m_2 =$ 1 TeV. 
Among all viable model points within each $h_2$ mass window we select 
the points which yield the maximum 
resonant di-Higgs production cross section [$(\sigma_{h_2}\times\mathrm{BR}_{h_2\to h_1 h_1})_\mathrm{max}$] (depending essentially on the values 
of $c_{\theta}$ and $\lambda_{211}$, as discussed in Section~\ref{Section_Constraints_Pheno}) to define 
twelve xSM strong first order EWPT motivated benchmarks.
This benchmark point set, which we refer to in the following as BM$^\mathrm{max}$, is presented in Table~\ref{T1}. 
Searches for resonant di-Higgs production in the $b \bar{b} W^+ W^-$ channel sensitive to this set of benchmarks
will be capable of probing into the strong first order EWPT region.
In the remainder of the paper we assess the LHC potential to probe such a strong first order EWPT via resonant di-Higgs production in the
$b \bar{b} W^+ W^-$ final state.

%


\section{Resonant Di-Higgs Production: the $b \bar{b} W^+ W^-$ Channel}
\label{sec:bbww}

As discussed above, in this work we explore the LHC sensitivity to resonant di-Higgs production in the 
$b \bar{b} W^+ W^-$ ($W^{+} \to \ell^{+} \nu_{\ell}$, $W^{-} \to \ell^{-} \bar{\nu}_{\ell}$) final state. By exploiting the two largest branching 
ratios of the 125 GeV Higgs boson $h_1$ we can retain sensitivity to smaller production cross sections, {\it i.e.}~larger $m_2$, and develop dedicated reconstruction 
approaches to suppress SM backgrounds. 
We require both $W$ bosons to decay leptonically (with $\ell = e,\,\mu$) to suppress the otherwise overwhelming
background from QCD multi-jet production.
The cancellation of momenta of two neutrinos in the $h_1 \to WW^* \to \ell\ell' \nu_{\ell} \bar{\nu}_{\ell'}$ decay
does not allow to reconstruct the invariant mass of the heavy resonance,
which substantially diminishes the LHC sensitivity to resonant di-Higgs production.    
To improve the sensitivity of the search, we develop a novel technique, called Heavy Mass Estimator,
designed to estimate the most likely invariant mass of the heavy $h_2$ state probabilistically.
The technique is conceptually similar to the Missing Mass Calculator 
(MMC) algorithm, which has previously been applied successfully to the mass
reconstruction of resonances decaying into $\tau^{+}\tau^{-}$ pairs~\cite{Elagin:2010aw,Spannowsky:2013qb}.

Throughout the remainder of the paper, we assume an LHC center-of-mass-energy of 13 TeV and
an integrated luminosity ranging between 300 fb$^{-1}$ and 3000 fb$^{-1}$, expected  to be collected between respectively the end of LHC Run 2 
data-taking (foreseen in 2022) and the end of the High Luminosity phase of LHC (foreseen in 2035).

\subsection{Monte Carlo Generation and Object Reconstruction}

For each of the xSM benchmark points in Table~\ref{T1} we generate our signal 
$p p \to h_1 h_1 \to  b \bar{b} W^+ W^-$ ($W^{+} \to \ell^{+} \nu_{\ell}$, $W^{-} \to \ell^{-} \bar{\nu}_{\ell}$)
using {\sc Herwig}++~\cite{Bahr:2008pv}.
The dominant SM background is top-pair ($t\bar{t}$) production\footnote[2]{Other potential (and largely subdominant) 
backgrounds such as Drell-Yan, diboson and single-top can be disregarded, as briefly discussed in Section~\ref{sec:selection}.}, 
which has been simulated at next-to-leading order (NLO) accuracy with {\sc Powheg}~\cite{Alioli:2010xd} and then subsequently processed 
with {\sc Herwig}++ for parton showering and hadronization to evaluate experimental sensitivity.
For simplicity, we restrict our signal and background Monte Carlo generation to $\ell = \mu$, 
but the subsequent sensitivity analysis takes into account the would-be contributions from final states 
with two electrons and one electron and one muon, for which we expect very similar efficiencies.


To evaluate the sensitivity achievable with the LHC data in this channel,
we use the CMS detector and performance parameters as a benchmark.
Assuming similar performance of the CMS and ATLAS detectors, in the combination we double the luminosity delivered per experiment.
The simulation of the CMS detector response is performed using {\sc Delphes} 3.3.0~\cite{deFavereau:2013fsa} and the recommended by CMS 
input card~\cite{deFavereau:2013fsa,CMS:1994hea}, with all 
reconstructed physical objects, such as tracks, calorimeter deposits, isolated muons, electrons, jets, and missing transverse energy $\MET$,
used in the data analysis.
Multiple proton-proton collisions during the same bunch-crossing (pileup) can have a strong impact on hadronic observables,
particularly during the high luminosity LHC runs, and we include in our reconstruction the effect of an average of 40 simultaneous proton-proton interactions.  
A particle-flow algorithm~\cite{deFavereau:2013fsa} has been successfully deployed in the CMS experiment and is implemented in {\sc Delphes} 
parametrically using the information from the tracking system and the calorimeters.
The particle-flow method is designed to reconstruct individual particles arising from collision by combining information
from relevant subdetectors, to improve the quality of particle identification and the performance of global event reconstruction.
Muons are reconstructed within the detector acceptance $|\eta_{\ell}|<2.4$.
Reconstruction and isolation selections follow the CMS definitions developed
for particle flow muons~\cite{Chatrchyan:2012xi} and use the medium working points for both.
Jets are reconstructed using the anti-$\text{k}_{t}$ ~jet algorithm~\cite{Cacciari:2008gp} with cone size R=0.4 by clustering the particle-flow tracks and particle-flow 
towers and we require $|\eta_j|<4.0$.
The jet area method~\cite{deFavereau:2013fsa} in {\sc Delphes} is applied in jet reconstruction 
to subtract the pileup contribution.
The b-tagging efficiency and mis-identification rates are modelled using the {\sc Delphes} parametrization of the CSV algorithm~\cite{Chatrchyan:2012jua}.
Tagging efficiencies and the mistag rates correspond to about 70\% and 1.5\%
for the medium working point and 85\% and 10\% for the loose working point, respectively.
The total transverse energy of a single event is calculated as the 2D-vector sum of the transverse momentum of all particles reconstructed by the CMS particle-flow algorithm.
The missing transverse energy is defined as the opposite of the total transverse energy, and it quantifies the transverse energy carried away from neutrinos.

\subsection{Invariant Mass Reconstruction for $h_2$: Heavy Mass Estimator}
\label{sec:mmc}

The cancellation of momenta of the two or more undetected neutrinos in the final state does not allow 
the reconstruction of the invariant mass of the  heavy scalar $h_2$ (and similarly for one of the 125 GeV scalars $h_1$)
using experimentally measurable quantities. 
To improve the analysis sensitivity, the HME technique,
a MMC-like probabilistic algorithm~\cite{Elagin:2010aw,Spannowsky:2013qb},
can be efficiently implemented  for the reconstruction of the mass of $h_2$.
To illustrate the implementation of the HME algorithm, 
we start with an idealized detector, in which properties of all visible particles are perfectly measured and
the missing transverse energy is equal to the negative vector sum of all visible particles.
The latter assumes that the missing transverse energy measurement is not affected by pileup.

We note that for the production process considered both 125 GeV $h_1$ states are on-shell, whereas one of the two $W$ bosons from the 
$h_1$ decay is typically off-shell (we use the label $1$ for the on-shell $W$, $W_1\rightarrow \mu_1 \nu_{\mu_1}$). 
With these simplified assumptions, the kinematics of the majority of the signal events satisfies the following:
\begin{eqnarray}
\MET_{x} = p_{x}(\nu_{\ell_1}) + p_{x}(\nu_{\ell_2})\label{ch4:1}\\
\MET_{y} = p_{y}(\nu_{\ell_1}) + p_{y}(\nu_{\ell_2})\label{ch4:2}\\
\sqrt{p^2(\ell_1,\nu_{\ell_1})} = M_W, \quad 20~\text{GeV} < \sqrt{p^2(\ell_2,\nu_{\ell_2})} < 45~\text{GeV}\label{ch4:3}\\
\left(p(\ell_1) + p(\ell_2) + p(\nu_{\ell_1}) + p(\nu_{\ell_2})\right)^2 = m_{h_1}^2\label{ch4:4}\\
(p(b_1) + p(b_2))^2 = m_{h_1}^{2}\label{ch4:5}
\end{eqnarray}
where $m_{h_1} \equiv m_1 = 125$ GeV, $\MET_{x}$, $\MET_{y}$ are the x- and y-components of the missing transverse 
energy $\MET$ vector, $p$ represent the various momentum 4-vectors
and $p_x$, $p_y$ are their x- and y-components. The momentum carried  by each neutrino is described
by three unknown momentum projections, leading to a total of five equations, one bound and six unknowns.

\begin{figure}[h!]
\centering
\includegraphics[width=0.485\textwidth]{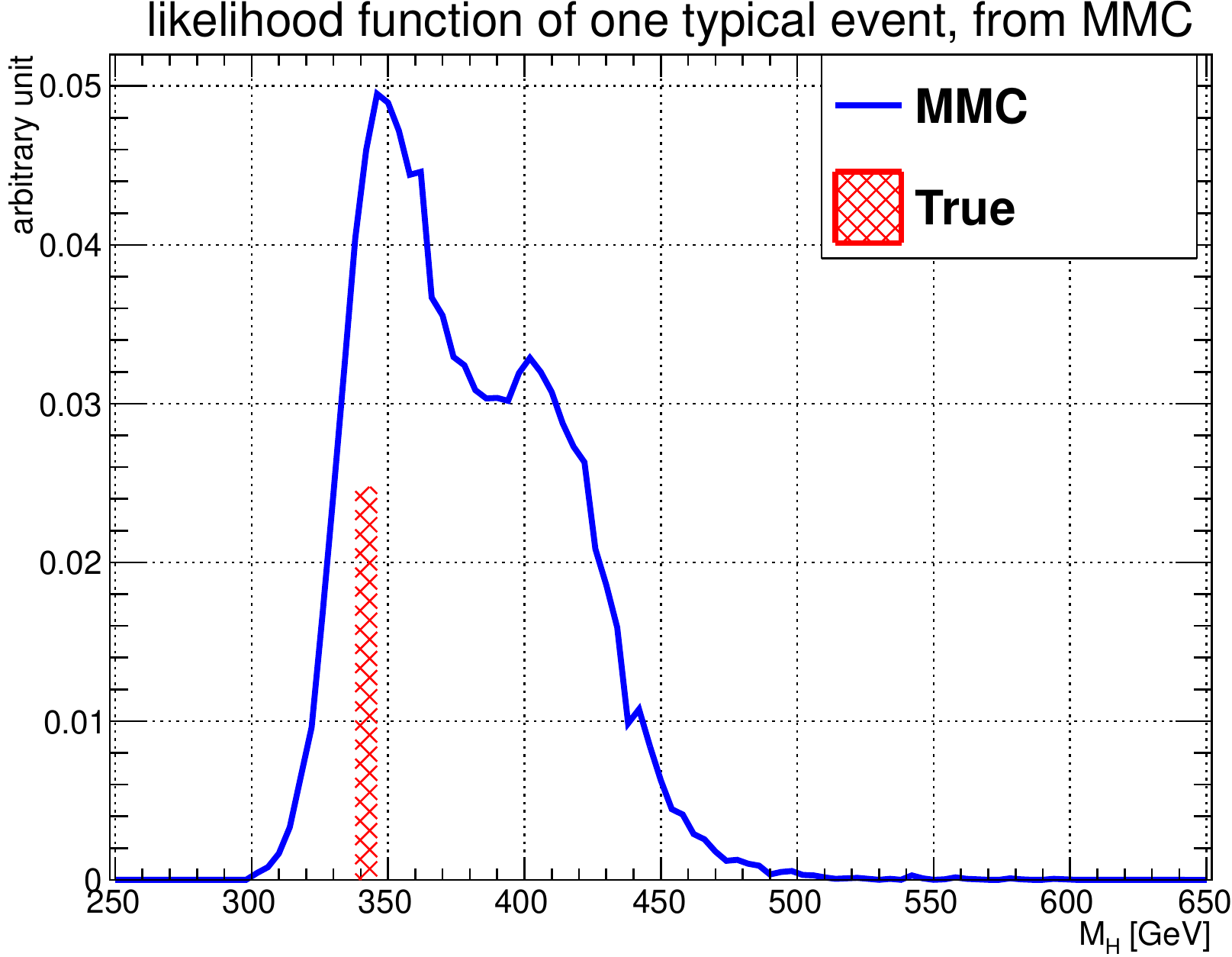} \hspace{2mm}
\includegraphics[width=0.485\textwidth]{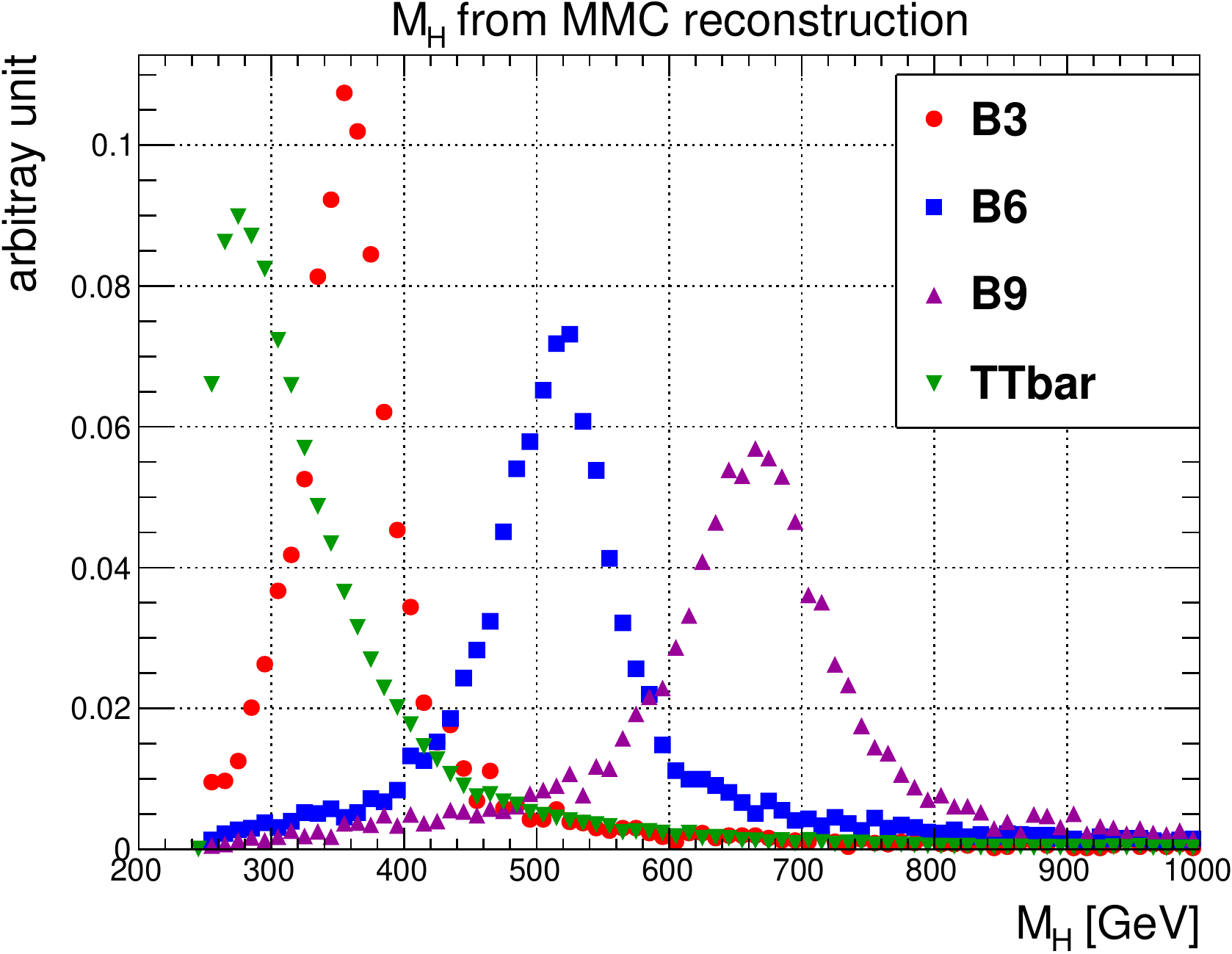}\\
\caption{\small Left: Global likelihood function (solid blue) computed from HME for a single signal event using the 
xSM benchmark point 3 (B3). The true value of the mass $m_2$ is marked by the red grid bar. Right:  HME distribution for B3 (red circle),
B6 (blue square), B9 (magenta up triangle) and $t\bar{t}$ (green down triangle). All distributions are normalized to unity.}
\label{bbww4:mmc}
\end{figure}

As seen from equations~\eqref{ch4:1}-\eqref{ch4:5}, four constraints 
reduce the number of unknowns to two, which we choose as the pseudo-rapidity $\eta_{\nu_1}$ and azimuthal angle $\phi_{\nu_1}$ of one neutrino.
Assigning random values to these two unknowns would then allow one to
scan the parameter space of allowed solutions to build a procedure to integrate over the
space of solutions consistent with the experimental measured quantities.
We refer to a single generation of the two unknowns as \textit{iteration}
if it respects the bound for the invariant mass of the off-shell W (else such single generation is discarded).
Each iteration yields an estimator for the mass of $h_2$: 
\begin{eqnarray}
m_2=(p(\ell_1) + p(\ell_2) + p(\nu_{\ell_1}) + p(\nu_{\ell_2}) + p(b_1) + p(b_2))^2 \label{ch4:6}
\end{eqnarray}
Furthermore, as not all pairs of values of the unknowns $\eta_{\nu_1}$ and $\phi_{\nu_1}$ are equally likely,
generating pairs of these values according to a suitably defined probability density function
would increase frequency of the estimated mass $m_2$ being close to the true value.
Such probability density function (PDF) can be obtained from a Monte Carlo simulation.
For each event, we generate thousands of iterations according to the PDF for $\eta_{\nu_1}$ and $\phi_{\nu_1}$,
and for each iteration we store the calculated value of $m_2$ building a probability distribution function for $m_2$, which we refer to as the 
HME \textit{global likelihood function.} 
In the full implementation of the algorithm, the values of $m_{h_1}$ and $M_W$ used in ~\eqref{ch4:3}-\eqref{ch4:5}
are generated according to Gaussian functions to account for the width of Higgs and $W$ bosons.
The addition of these two \textit{variables} effectively increases the dimensionality of the space in which the scan is performed to four.
The introduction of additional probability density functions to account for realistic resolutions
of experimental measurements further increases the dimensionality of the space scanned.
One of the most essential additions accounts for the b-jet energy mismeasurements
which leads to the invariant mass of the two b-jets being on average lower than $m_{h_1}$.
We compute and apply an energy correction extracted from the simulation for the leading b-jet,
and use equation~\eqref{ch4:5} to correct the energy of the sub-leading b-jet.
This procedure simultaneously improves the missing transverse energy estimation
used in ~\eqref{ch4:1}-\eqref{ch4:2}, that is finally smeared according to the detector resolution predicted by {\sc Delphes}.
Fig.~\ref{bbww4:mmc} (left) provides an illustration of a typical HME global likelihood for a single event, which
peaks near the true value of the heavy scalar mass $m_2$.
In this analysis, we use the most probable mass from the likelihood as the estimator of the heavy Higgs mass $m_2$.
Note that selecting the peak position of a single event global likelihood as the estimator is the simplest solution,
which one could likely improve upon by utilizing more information on the shape of the likelihood or even using the entire distribution in the analysis.
Fig.~\ref{bbww4:mmc} (right) shows the reconstructed $m_2$ mass for various xSM benchmark scenarios described in Table~\ref{T1}.


\vspace{2mm}

\subsection{Analysis Selection}
\label{sec:selection}


The experimental signature consists of two energetic leptons, two energetic b-tagged jets and significant missing transverse energy due to neutrinos. 
As discussed above, the dominant SM background process is $t\bar{t}$, 
with a very large production cross section (see {\it e.g.}~\cite{CMS-PAS-TOP-15-010}).
Other potential SM backgrounds are Drell-Yan, single-top, di-boson and $t\bar{t}V$ production~\cite{CMS:2016rec}, as well as production of the SM Higgs boson
(decaying to $W W$) in association with jets ({\it e.g.} in vector boson fusion).
However, it has been shown in~\cite{Martin-Lozano:2015dja} 
(see also~\cite{CMS:2016rec}) that basic selection criteria together with mild kinematic cuts on $p_T(b \bar{b})$, $p_T(\ell\ell)$
and the invariant mass of the $b \bar{b}$ and di-lepton systems yield all these other backgrounds to be negligible,
while maintaining a high signal efficiency.
We therefore can safely disregard them in the present work.

\begin{table}[h!]
 \begin{center}
 \caption{Pre-MVA selection.}
 \label{tab:selectioncuts}
 \begin{tabular}{cl}
   \hline
   Variable  & Cut \\
   \hline
   \hline 
   $\Delta R(\ell\ell)$  & $ 0.07<\Delta R(\ell\ell)<3.3$\\
   $\Delta R(jj)$  & $\Delta R(jj)<5.0$ \\
   $m(\ell\ell)$            & $5$ GeV $<m(\ell\ell)< 100$ GeV \\
   $m(jj)$            & $m(jj) > 22$ GeV \\
   \hline
 \end{tabular}
 \end{center}
\end{table}

Initial event pre-selection is performed as follows: we require the presence of two muons\footnote[3]{We recall that we have restricted our analysis to muons, 
yet it will apply equally to $\ell = e, \mu$.} 
with opposite sign and $p_T \geq 10$ GeV, $|\eta|<2.4$; if more than two muons are present in the event,
the two oppositely charged muons with largest transverse momentum are selected.
In addition, at least two b-tagged jets with $p_T > 30$ GeV and $|\eta|<2.5$ are required.
At least one of the two b-jet candidates has to be b-tagged using the CSV algorithm at the medium working point,
while the other jet is only required to satisfy at least the loose b-jet requirement.
If more than two b-jets satisfy all selection criteria, the two b-jets candidate with the invariant mass closer to $m_{h_1} = 125$ GeV are selected.
Finally we require the missing transverse energy to be $\MET > 20$ GeV.
After event pre-selection, we also perform a set of kinematic cuts (pre-MVA selection) summarized in Table~\ref{tab:selectioncuts}, 
which reject approximately 5\% of the signal events (for all signal mass points) and about 40\% of $t\bar{t}$ events.

\begin{figure}[h!]
\centering
\includegraphics[width=0.325\textwidth]{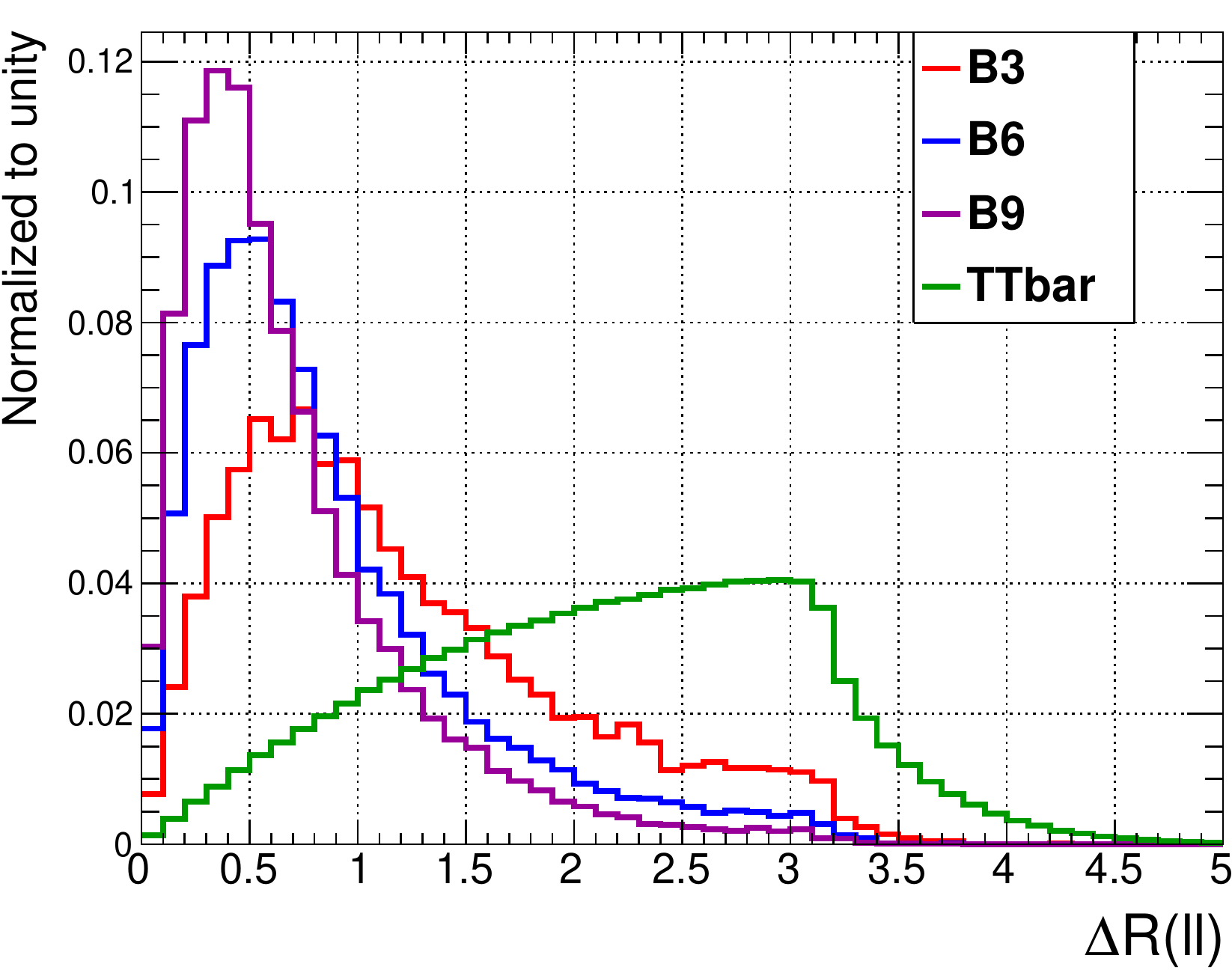}
\includegraphics[width=0.325\textwidth]{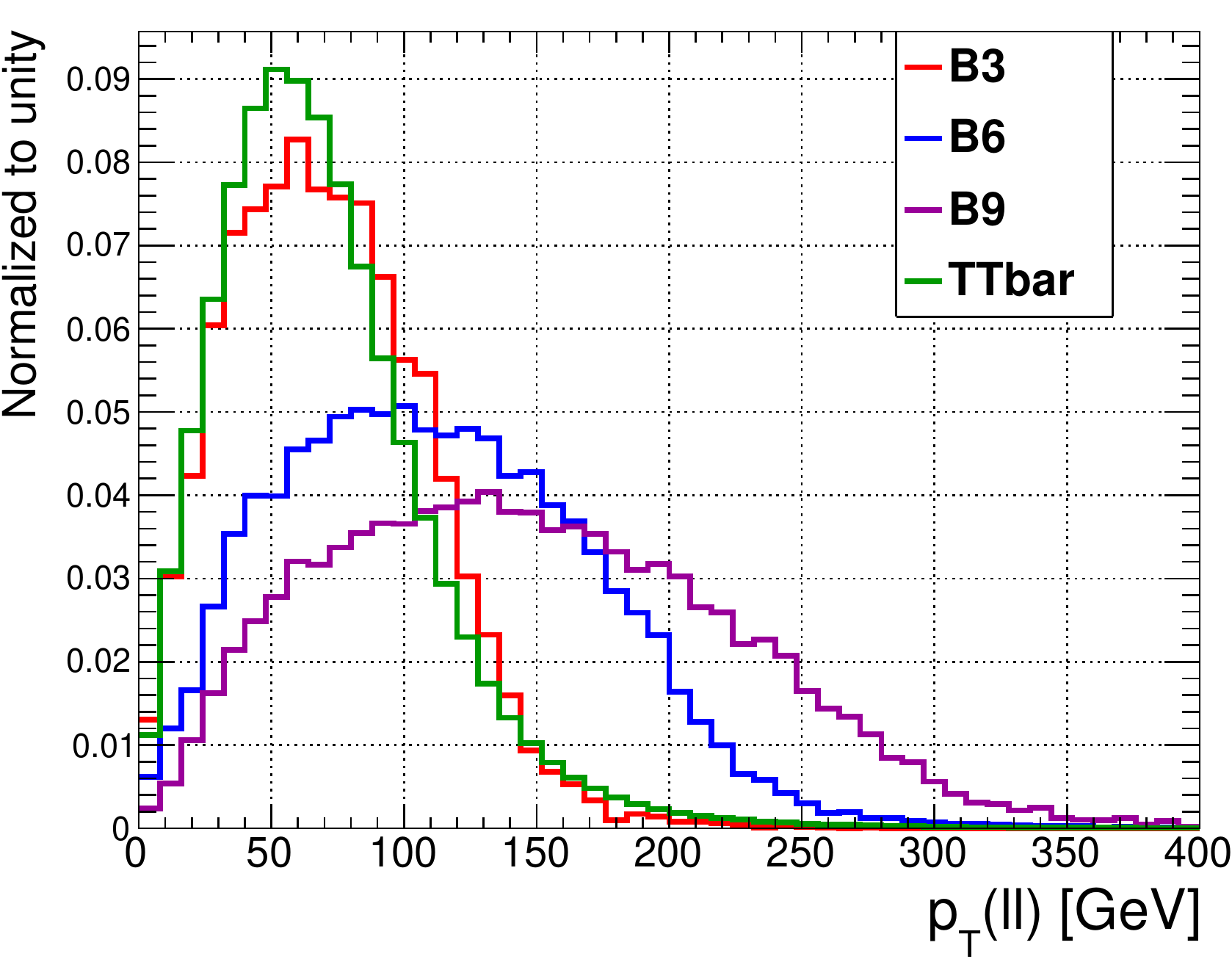}
\includegraphics[width=0.325\textwidth]{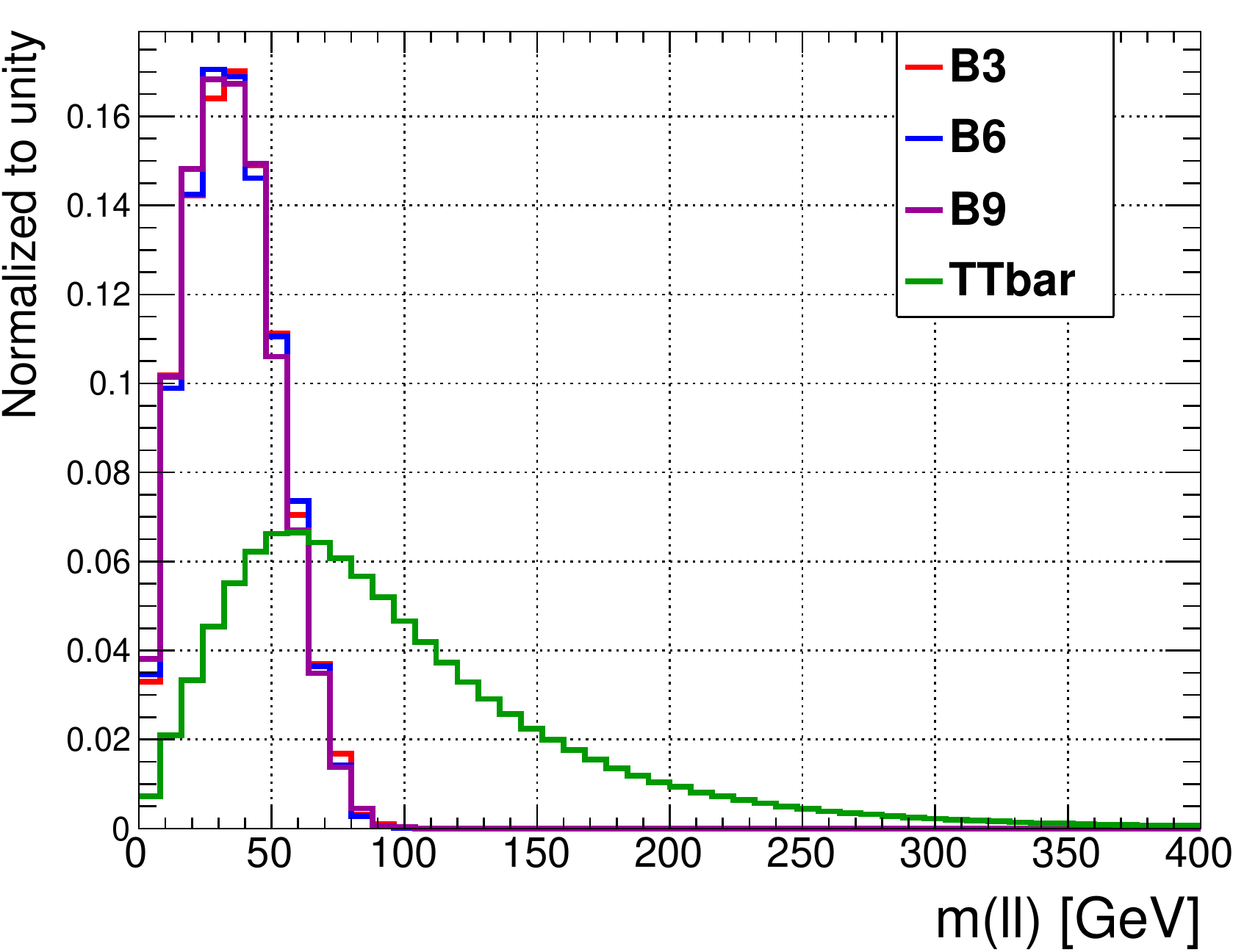}

\includegraphics[width=0.325\textwidth]{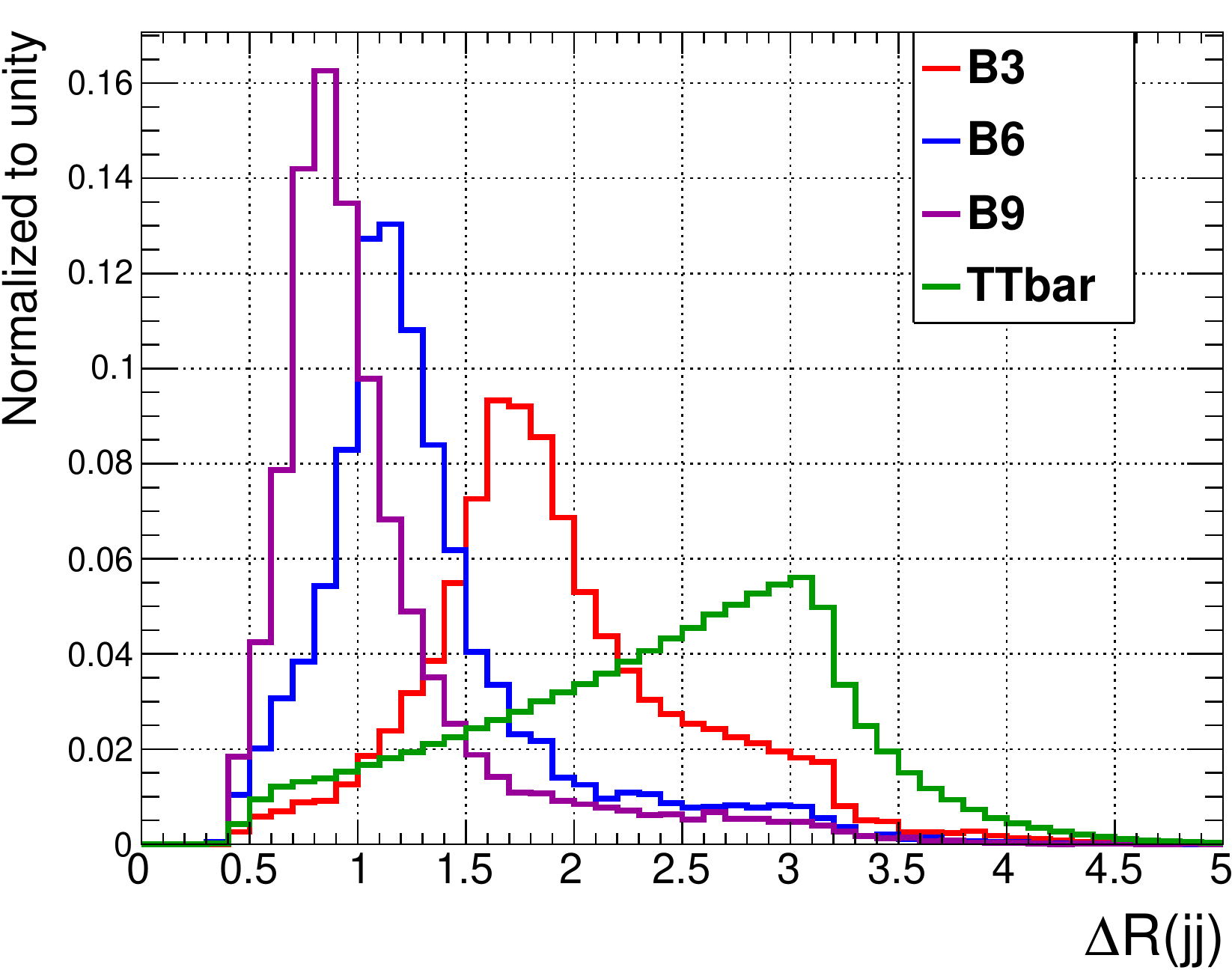}
\includegraphics[width=0.325\textwidth]{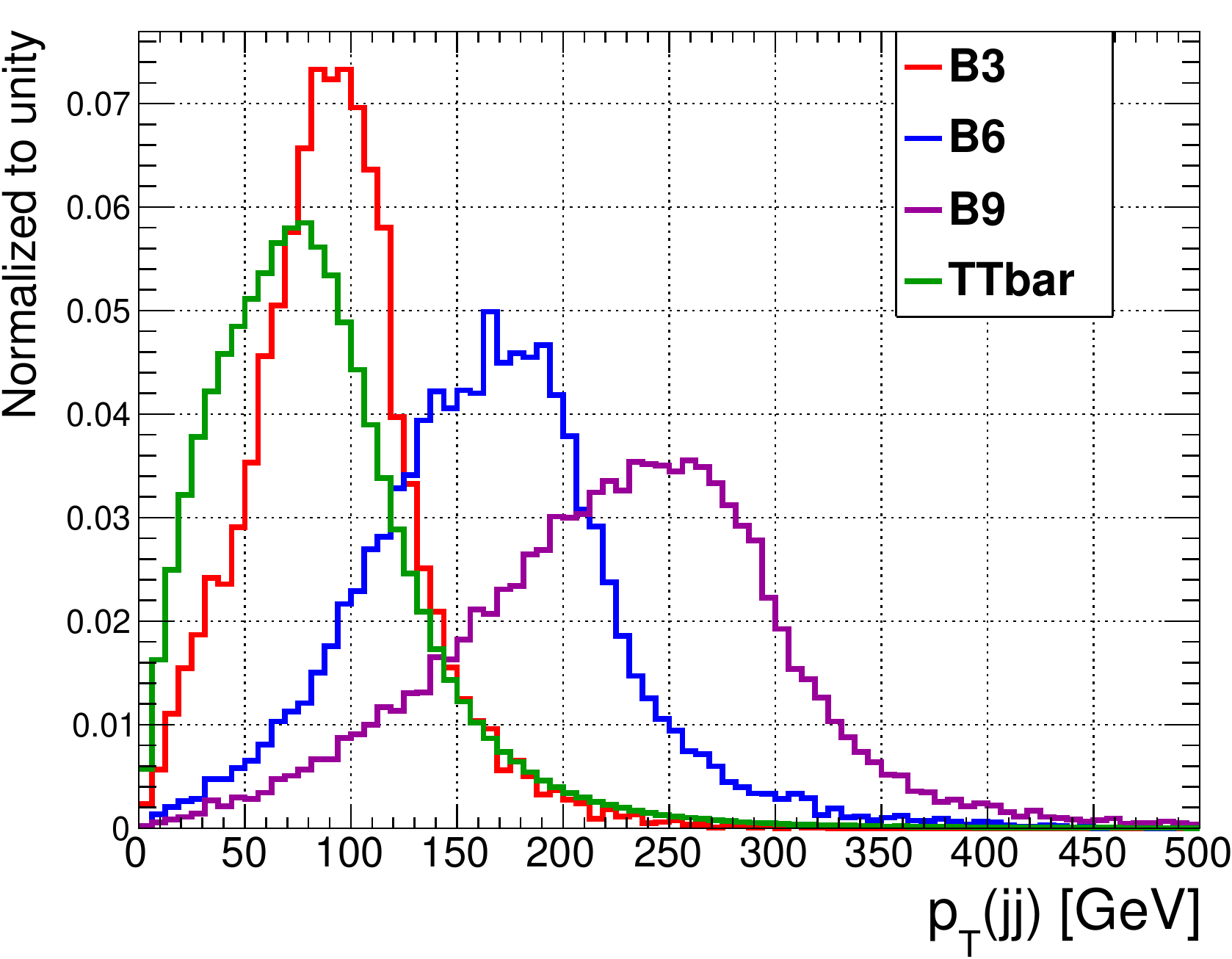}
\includegraphics[width=0.325\textwidth]{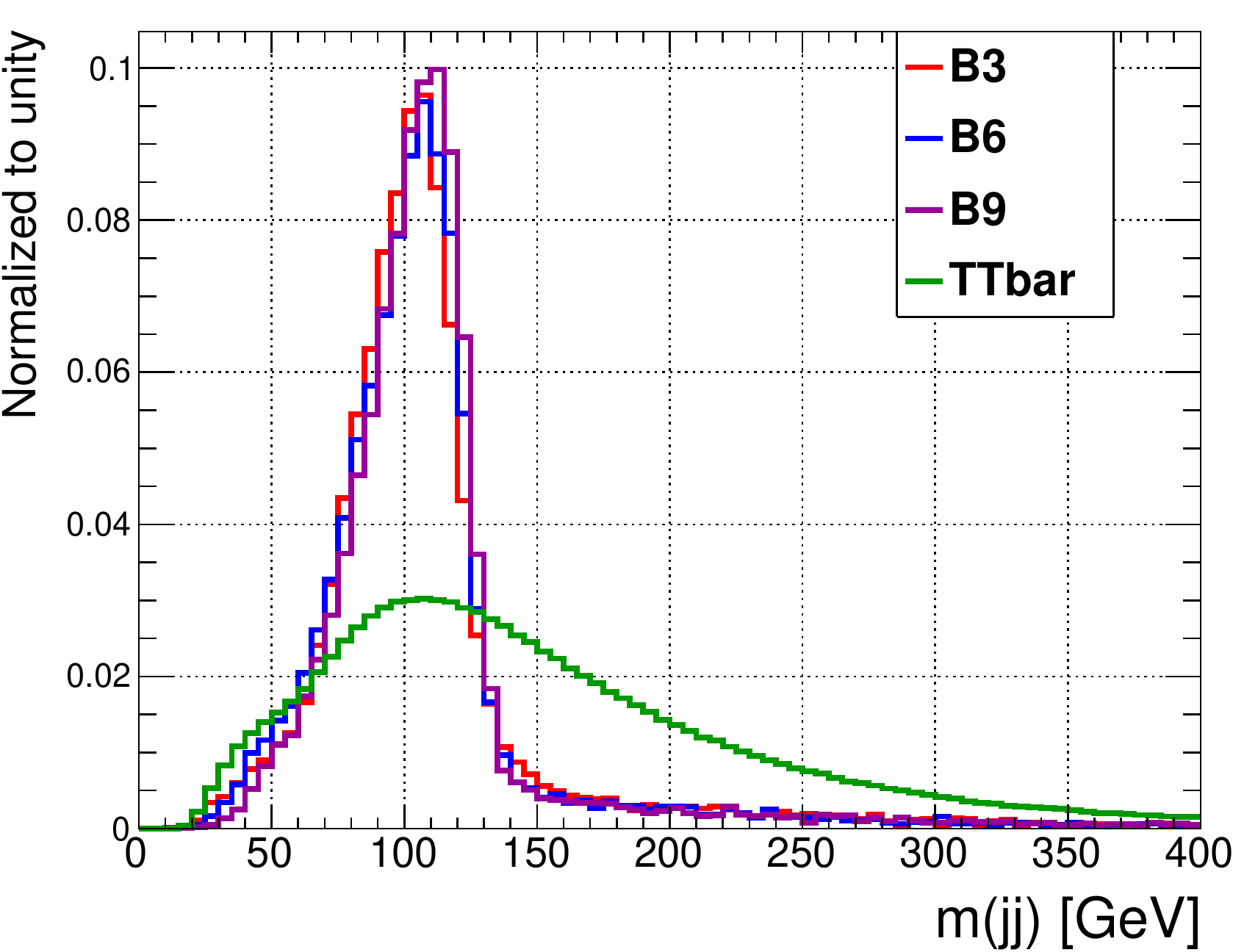}
\caption{\small Kinematic variables $\Delta R(\ell\ell)$ (top left), $p_T(\ell\ell)$ (top middle), $m(\ell\ell)$ (top right), $\Delta R(jj)$ (bottom left),
$p_T(jj)$ (bottom middle), $m(jj)$ (bottom right), with distributions normalized to unity.}
\label{bbWW1:kinematics}
\end{figure}

\begin{figure}[h!]
\centering
\includegraphics[width=0.485\textwidth]{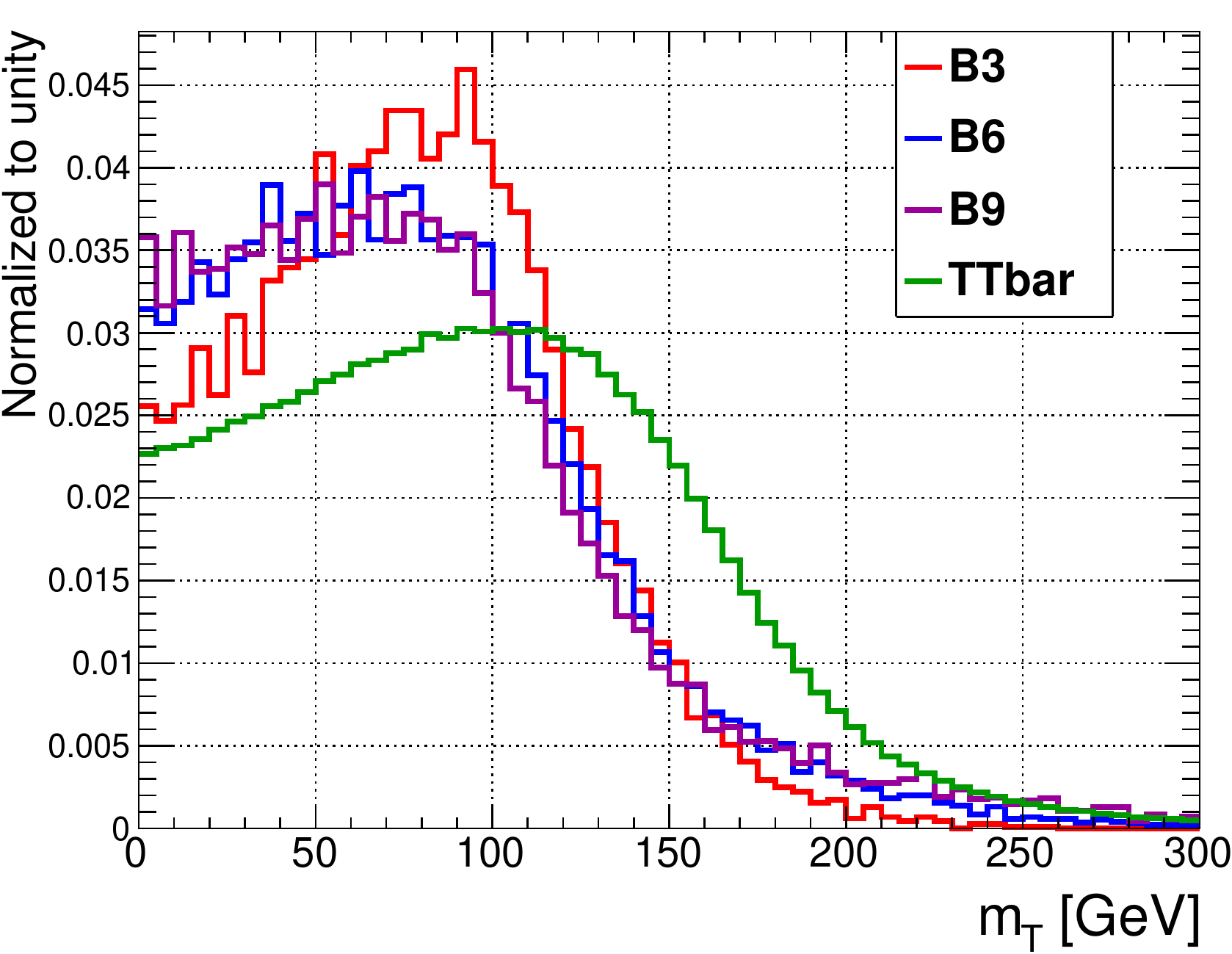} \hspace{1mm}
\includegraphics[width=0.485\textwidth]{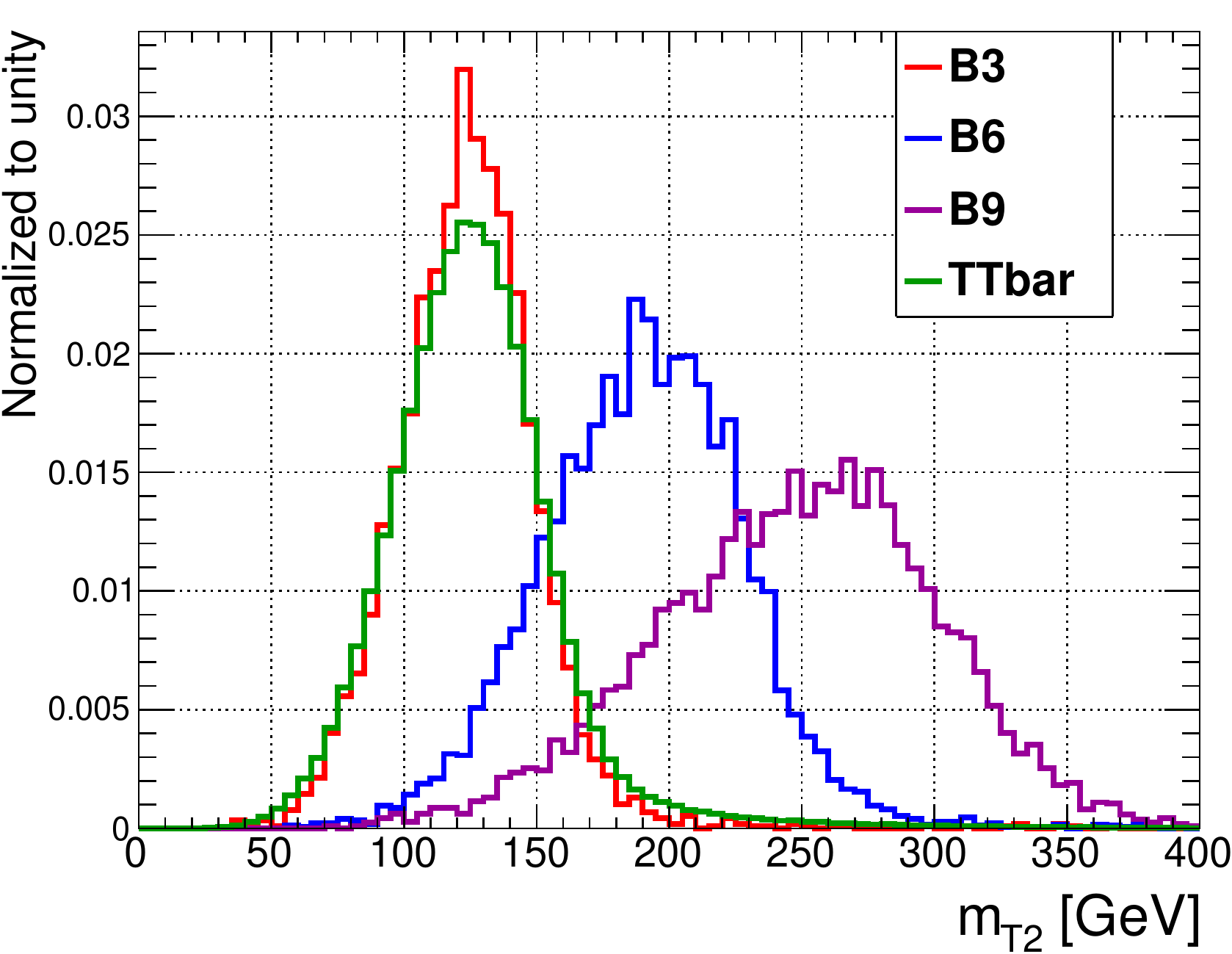}
\caption{$m_{T}$ (left) and $m_{T2}$ (right) distributions after the pre-MVA selection, with distributions normalized to unity.}
\label{bbWW1:kinematics2}
\end{figure}

In order to optimize the background and signal discrimination
we have tested several MVA algorithms (Likelihood, LikelihoodMIX, KNN, MLP, BDT, and BDTD) 
available from the TMVA package (version 4.1.2)~\cite{Hocker:2007ht}, choosing 
the algorithm most performing in terms of background/signal discrimination as a function of $m_2$:
a BDT for low mass (xSM benchmarks B1 to B7), and a Likelihood method for high mass (xSM benchmarks B8 to B12).
The training of the MVA has been done independently for each signal mass point\footnote[4]{The use 
of discrete mass values in this work is a simplification; in the actual data analysis, training of the MVA would 
be performed to optimize sensitivity within ranges of \textit{target masses} $m_2$. Effectively, this would split the analysis 
in several sub-analyses, each optimized for a specific range of target masses $m_2$.
}
considered in the analysis using the discriminating variables
$\Delta R(\ell\ell)$, $p_T(\ell\ell)$, $m(\ell\ell)$, $\Delta R(jj)$, $p_T(jj)$, $m(j j)$,
$\Delta R(\ell,j)$, $\Delta R(\ell\ell,jj)$, $\Delta R_{\mathrm{min}}(\ell,j)$, 
$\Delta \phi(\ell\ell,jj)$, $m_T$ and $m_{T2}$. 
The variable $\Delta R_{\mathrm{min}}(\ell,j)$ is computed by measuring the $\Delta R$ between each lepton and each jet
and selecting the smaller among these values.
Kinematic distributions after pre-selection for the first six variables above are shown in Fig.~\ref{bbWW1:kinematics} 
for the xSM signal samples B3, B6 and B9 together with the $t \bar{t}$ background. 
The transverse mass variable $m_T$ is defined as
\begin{equation}
 m_T = \sqrt{2\, p_T(\ell \ell) \MET \left[ 1-\cos(\phi_{\ell\ell}-\phi_{\MET})\right]}\, .
\end{equation}
The $m_{T2}$ variable~\cite{Lester:1999tx,Barr:2003rg} provides a transverse mass estimate in systems where more than one neutrino is present, treating the lepton 
and the b-jet as a single object. Fig.~\ref{bbWW1:kinematics2} shows the $m_T$ and $m_{T2}$ distributions (normalized to unity) after the pre-MVA selection.
The discriminating power of the $m_{T2}$ variable in di-Higgs final states, already appreciated in~\cite{Barr:2013tda}, is good for all signal samples where the invariant mass 
of the heavy resonance is greater than twice the top mass.

Table~\ref{tab:EF} shows the expected event yield $N_{\mathrm{event}}$ with 300 fb$^{-1}$ of integrated luminosity 
for the $t\bar{t}$ background and each of the xSM signal benchmarks after pre-selection and pre-MVA selection cuts,
as well as the final yield of signal ($N_S$) and background ($N_B$) events after the MVA selection in each benchmark scenario.
%
%
After the MVA-based selection is applied, the invariant mass of the heavy Higgs scalar $h_2$ is reconstructed
for the surviving events using the HME probabilistic technique described in Section~\ref{sec:mmc}.
The HME distribution for signal and background is then used for setting upper limits on the signal production cross section.

\begin{table}[h]

 \caption{Number of signal ($N_S$) and background ($N_B$) events expected collecting 300 fb$^{-1}$ of integrated luminosity
after applying different stages of selection and prior to the final fit using the HME mass estimator distribution.}
 \label{tab:EF}
  \resizebox{\columnwidth}{!}{%
 \vspace{2mm}
 
\centering
 \begin{tabular}{c|c|c|c|c}

   \hline
   Process  & $N_{\mathrm{event}}$ (pre-selection) & $N_{\mathrm{event}}$ (pre-MVA cuts) & $N_{S}$ (MVA) & $N_{B}$ (MVA)\\
   \hline
   \hline 
   B1    &395 & 383  & 183 & 6962\\
   B2    &395 & 385  & 171 & 27372\\
   B3    &318 & 310  & 152 & 26593\\
   B4    &137 & 134  & 35 & 1425\\
   B5    &104 & 102   & 19 & 193\\
   B6    &52   & 50     & 12 & 95\\
   B7    &31    & 30    & 10 & 91\\
   B8    &22    & 21    & 4.5 & 28\\
   B9    &13    & 12    & 3.2 & 23\\
   B10  &5     & 5       & 1.2 & 13\\
   B11  &3     & 3       & 0.8 & 10\\
   B12  &1     & 1       & 0.2 & 4.5\\
   $t\bar{t}$ Background &782721 &382836   &  &  \\
   \hline
 \end{tabular}
 }
\end{table}

%

\subsection{Systematic Uncertainties}
\label{sec:syst}
For the systematic uncertainties in evaluating signal acceptance we assume, based on previous publications presented by the CMS collaboration,
a conservative systematic uncertainty of about 10\%~\cite{4muPap,ChargHiggs,bbgg,bblow}.
This systematic includes the uncertainty on the integrated luminosity,
the uncertainty on the trigger efficiency, the lepton identification and isolation,
the uncertainty on the parton distribution functions ({\sl PDFs}) and on the factorization and renormalization scales.

The systematic uncertainties associated with the precision in the knowledge of the background shape and normalization
can significantly affect the sensitivity of the analysis.
In the absence of real data, we illustrate a possible way to estimate these uncertainties,
which we believe to be conservative, as the experimentalists are likely to deploy
more sophisticated approaches to reduce the impact of systematic uncertainty on the sensitivity
and we also anticipate improvements in the quality of the description of the $t\bar{t}$ background
arising from theoretical efforts and Monte Carlo tuning.
For the purposes of this study, we define a control region dominated by the background,
and use it to compare data and simulation to obtain a scale-factor (SF) for correcting the simulation
prediction for the $t\bar{t}$ contribution.
The control region is designed to contain background events with properties and kinematics as in the main signal region.

Since the $t\bar{t}$ kinematics has no strong dependence on the di-jet or di-lepton invariant mass,
we define control regions by selecting events with the measured di-jet invariant mass greater than 150 GeV
or di-lepton invariant mass greater than 100 GeV (see Fig.~\ref{bbWW1:kinematics}).
Using more than one control region allows to cross-check and validate the scale-factors
and, if needed, adjust the uncertainty associated with the scale factors.
Once the control region is defined we apply the same kinematic selections and perform the MVA training exactly as it is done
for the main analysis.
We choose the MVA cut that yields the same background rejection as the cut that has been found
to yield optimal sensitivity for the same target mass point in the main analysis.
For all mass points, the signal contribution remains negligible in the control regions.
Finally, the yield of surviving background events is used to derive the uncertainty in the scale factor
(in our case, the same events play the role of both the "data" and the "prediction" so the mean value of the scale factor is by definition equal to unity).
Following this methodology, we estimate the systematic uncertainty on the knowledge of the background normalization to be 1\% for the signal samples B1, B2 and B3;
5\% for B4; 10\% for B5; 12\% for B6, B7, B8, B9, B10; and  15\% for B11 and B12.
In the final sensitivity estimates, for the three lowest mass points we chose to increase the systematic uncertainty
for the background normalization from 1\% to 3\%. This has been driven by the considerations that the lower
mass ranges are the most susceptible to the knowledge of the background normalization
(this is because the EHM mass for $t\bar{t}$ and the lower mass signal samples are the most alike).
Furthermore, the kinematic phase space of the control region never fully emulates the phase space of the signal region
and so actual data analyses are likely to use several control regions to ensure good control of the background normalization,
which is likely to increase the systematic uncertainty.

\section{Prospects for LHC Run 2 and HL-LHC}
\label{sec:sensit}

Once the full set of selections is applied to the signal and background samples, and the systematic uncertainties are defined,
we compute the expected limits on the resonant di-Higgs production cross section multiplied
by the $h_2 \rightarrow h_1h_1$ branching fraction ($\sigma \times \mathrm{BR}$).

\begin{figure}[h!]
\centering
\includegraphics[width=0.49\textwidth]{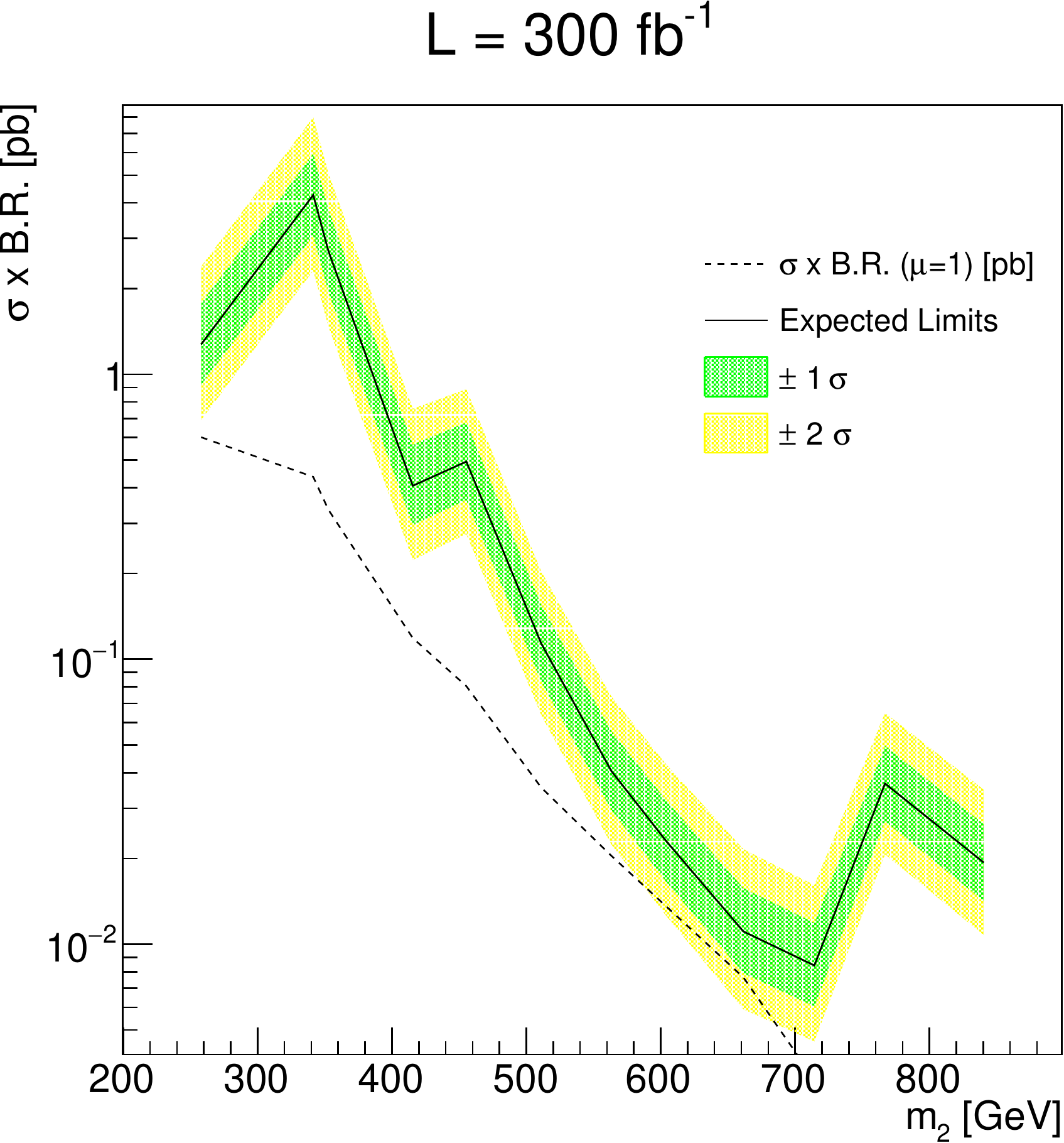}\vspace{1mm}
\includegraphics[width=0.49\textwidth]{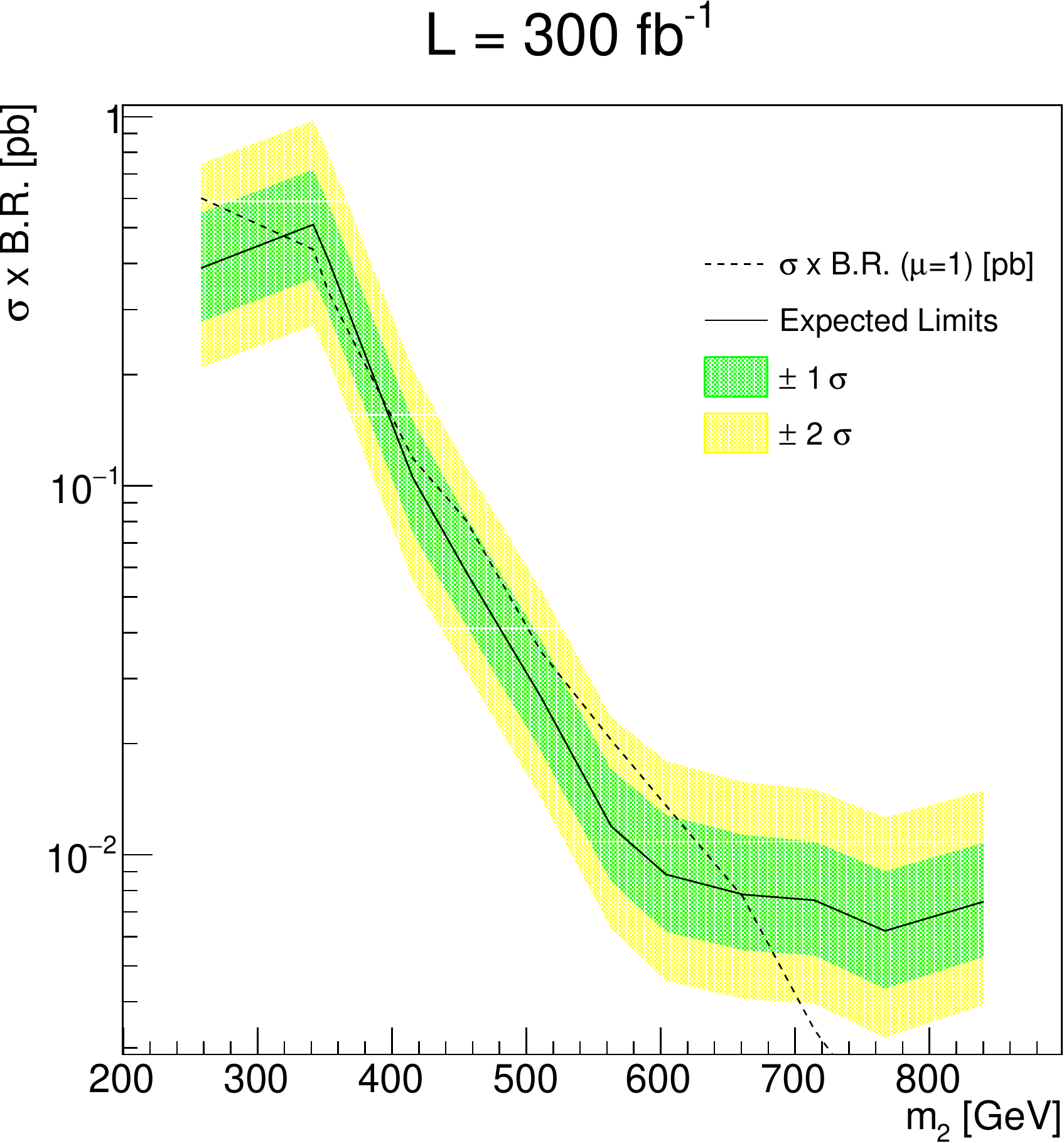}
\caption{The dashed line represent the cross section times $h_2 \rightarrow h_1h_1$ branching ratio for each mass sample in the set BM$^\mathrm{max}$.
The continuous black line instead represents the predicted 95\% C.L. upper limit on $\sigma \times \mathrm{BR}$
with 300 fb$^{-1}$ of data collected.
On the left  the limits are obtained performing a cut-and-count experiment on the final number of signal and background events.
The limit in this case is entirely driven by the high yield of the $t\bar{t}$ background events and the
discontinuities are due to the non-continuous assigned systematic uncertainty of the scale factors for the background normalization.
The confidence intervals for the expected limit are given at 68\% and 95\% coverage probability.
On the right limits are computed by fitting the HME distribution.
}
\label{limits_mmcorno}
\end{figure}

\begin{figure}[h!]
\centering
\includegraphics[width=0.65\textwidth]{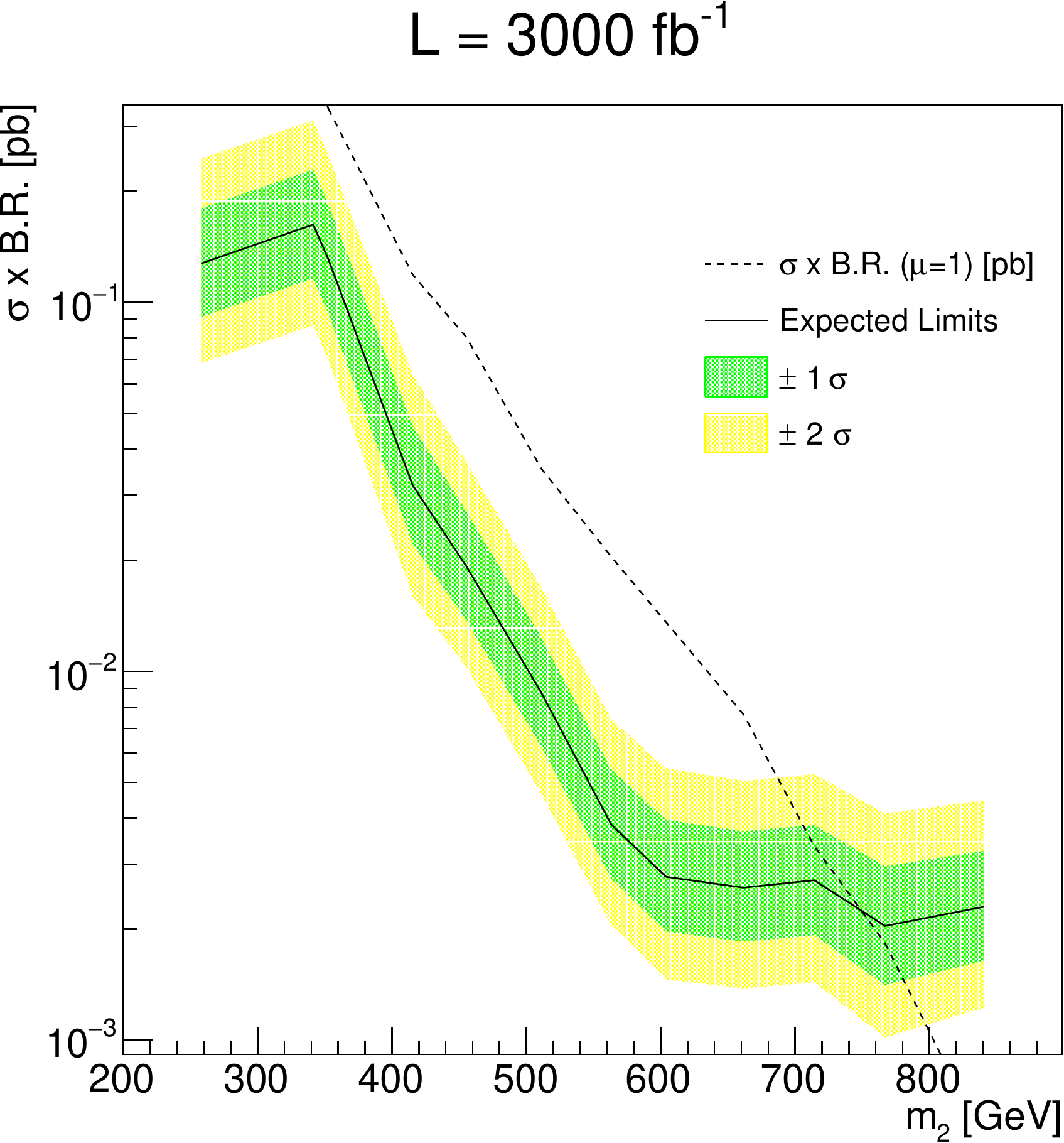}\\
\caption{The dashed line represent the cross section times $h_2 \rightarrow h_1h_1$ branching ratio for each mass sample in the set BM$^\mathrm{max}$.
The continuous black line instead represents the $\sigma \times \mathrm{BR}$ excluded at 95\% C.L. in case 3000 fb$^{-1}$ of data collected.
The confidence intervals for the expected limit are given at 68\% and 95\% coverage probability.
}
\label{limits}
\end{figure}

For calculations of expected limits, shown in Fig.~\ref{limits_mmcorno}, we adopt the modified frequentist criterion CLs~\cite{bib:limi_83}. 
The chosen test statistic, used to determine how signal- or background-like the data are, is based on the profile likelihood ratio~\cite{bib:limi_84}.
Systematic uncertainties are incorporated in the analysis via nuisance parameters and are treated according to the frequentist paradigm. 
Results presented in this paper are obtained using asymptotic formulae~\cite{asymp_101140}.
The dashed line represents the cross section times branching fraction expected for each mass point within BM$^\mathrm{max}$
(note that benchmark models are chosen as models yielding highest cross-section
for each mass range individually, which affects the smoothness of the theoretical prediction curve).
The continuous black line represents the predicted 95\% C.L. upper limit on the
cross-section times branching fraction $\sigma_{h_2}\times\mathrm{BR}_{h_2\to h_1h_1}$.
In Fig.~\ref{limits_mmcorno} (left) the limits are shown using a cut-and-count analysis, {\em i.e.}
applying the whole selection and counting the number of signal and background events expected at the end of the analysis.
In this scenario the Heavy Mass Estimator developed in this study has not been used.
This simplified method has substantially higher background contamination,
with the limit being entirely driven by the background level.
The discontinuity in the limit is an artefact of using a non-continuous systematic uncertainties
in the scale factors for background normalization (remember that the limits are calculated
only for the discrete set of mass points connected by a line to guide reader's eye).
In Fig.~\ref{limits_mmcorno} (right) the upper limits are computed by fitting the HME distribution,
in which case the background under the signal peak is substantially lower and the use of
the fit procedure reduces dependence on the background normalization scale factors.
Limits are computed assuming a search with 300 fb$^{-1}$ of integrated luminosity,
using both electrons and muons in the final state, and presuming an eventual combination of the ATLAS and CMS data.
Weakening of the limit at $m_2 \sim$ 350 GeV is due to the similarity of the HME shapes
for the $t\bar{t}$ background and signal in that mass range.
Above $m_2 \sim$ 600 GeV the limit trend changes.
In this region, the number of $t\bar{t}$ events in the signal region becomes almost constant,
as can be seen in the $t\bar{t}$ HME distribution in Fig.~\ref{bbww4:mmc} right, and the limit follows the same trend.

\begin{figure}[h!]
\centering
\includegraphics[width=0.65\textwidth]{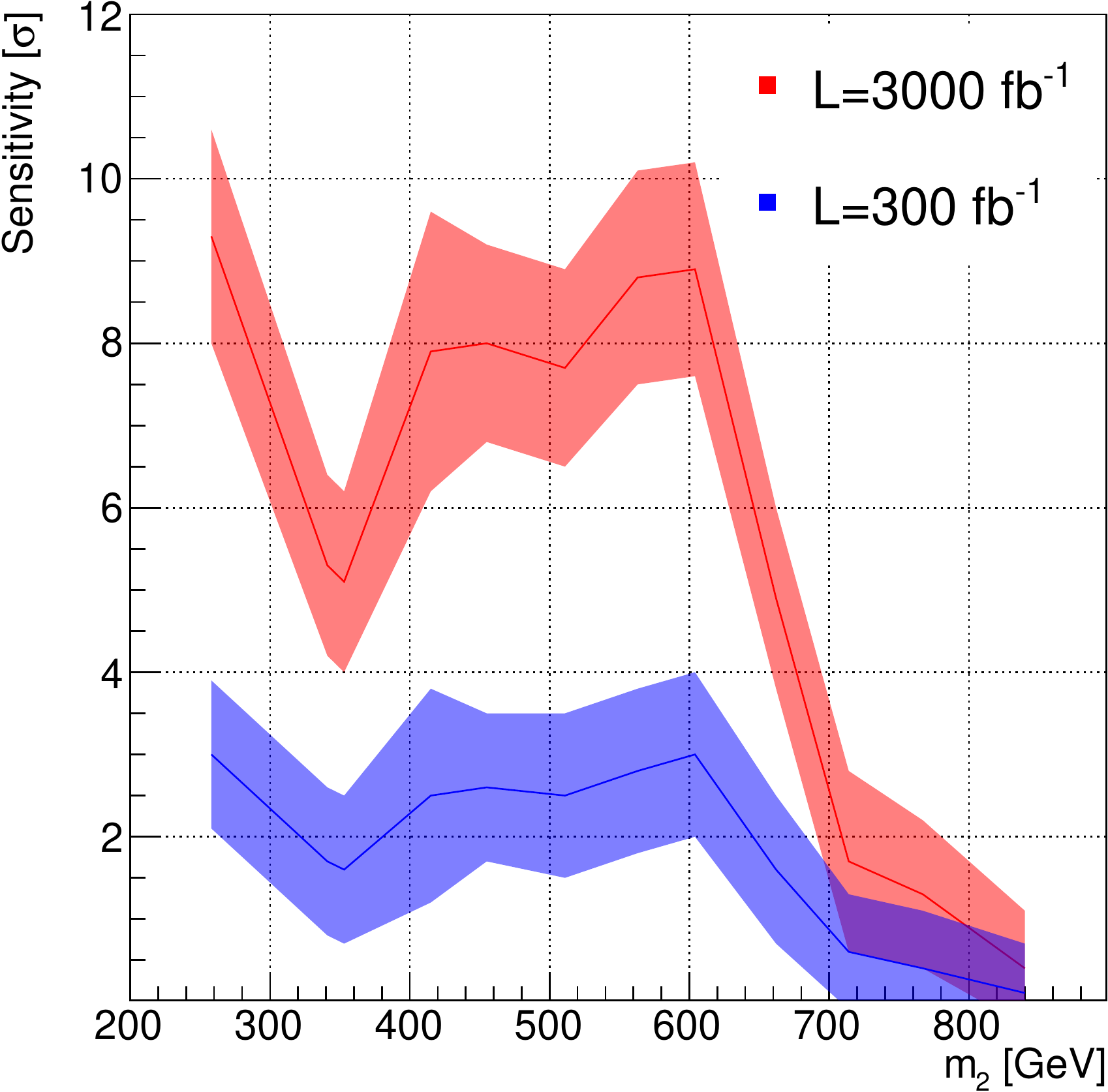}
\caption{
The colored central lines represent the sensitivity of the analysis assuming 300 fb$^{-1}$ of integrated luminosity (blue curve) and 3000 fb$^{-1}$ (red curve).
The confidence interval on the sensitivity central value is given at 68\% coverage probability.}
\label{sensitivity}
\end{figure}

In Fig.~\ref{limits} the limits are shown assuming 3000 fb$^{-1}$ of integrated luminosity.
We obtained such limits by scaling the number of event in signal and background. We did not
re-simulate the pileup scenario. 
The local expected significance of the analysis is computed generating
toy-models following the background hypothesis and the same
profile likelihood ratio-based CLs technique that has been used for deriving the limits.
The local p-value is then converted into significance $\sigma$ presented in Fig.~\ref{sensitivity}.
Both the limits and the sensitivity correspond to a combination of $ee$, $\mu\mu$ and $e\mu$ channels
with signal and background selection efficiencies equal to those for the $\mu\mu$ channel
studied in this paper, and an eventual combination of the CMS and ATLAS results.
The sensitivity is shown assuming 300 fb$^{-1}$ (blue curve) and 3000 fb$^{-1}$ (red curve)
of integrated luminosity per experiment.
We find that 3000 fb$^{-1}$ of data
could potentially be sufficient for a potential discovery for $m_2\lesssim 700$ GeV,
possibly with the exception of the region around $m_2$ = 350 GeV where fluctuations and the
look-elsewhere effect may bring the global significance below the conventional 5$\sigma$ threshold.
The confidence interval on the sensitivity central value is given at 68\% coverage probability.

\begin{figure}[h!]
\centering
\includegraphics[width=0.7\textwidth]{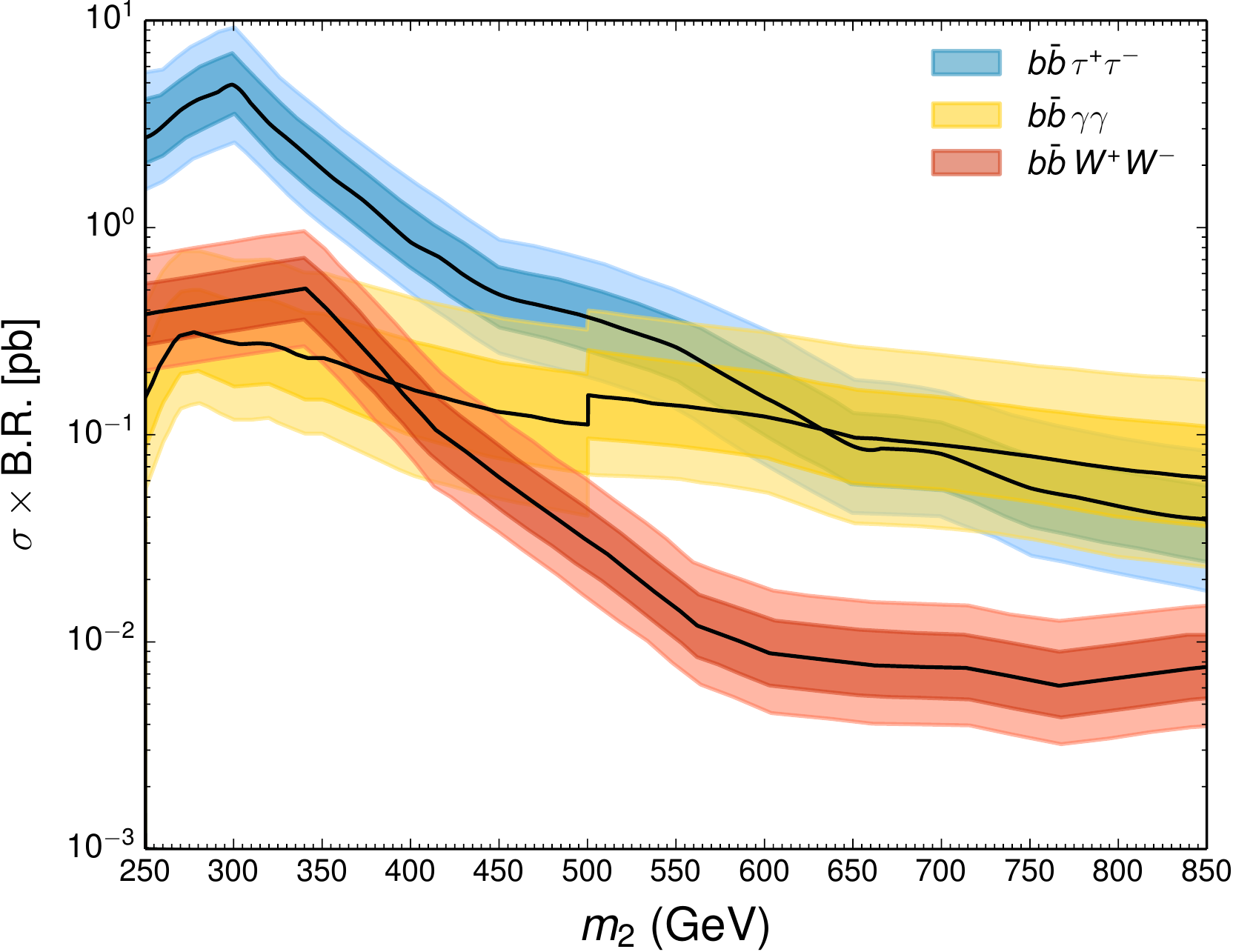}
\caption{13 TeV LHC projected 95\% C.L. limits (solid-black lines) on $\sigma_{p p \to h_2} \times \mathrm{BR}_{h_2 \to h_1 h_1}$ (in pb) for an integrated luminosity 
$\mathcal{L} = 300$ fb$^{-1}$ and assuming an ATLAS-CMS combination, in the $b\bar{b} W^{+}W^{-}$ final state (as shown in Fig.~\ref{limits_mmcorno}, Right) 
and in the $b \bar{b} \tau^{+} \tau^{-}$ and $b \bar{b} \gamma \gamma$ final states 
(through a na\"ive $\sqrt{\mathcal{L}}$ extrapolation of the resonant di-Higgs 13 TeV CMS analysis in the $b \bar{b} \tau^{+} \tau^{-}$~\cite{CMS:2016knm} and 
$b \bar{b} \gamma \gamma$~\cite{CMS:2016vpz} final states). In all cases the dark (pale) colored bands correspond to the confidence intervals for 
the expected limit at 68\% (95\%) coverage probability.}
\label{limbbtautau}
\end{figure}

Finally, we compare the sensitivity of the $b \bar{b} W^{+} W^{-}$ channel with that of the $b \bar{b} \tau^{+} \tau^{-}$ and $b \bar{b} \gamma \gamma$ channels.
We extrapolate the current, public 13 TeV limits from the CMS search for resonant di-Higgs in $b \bar{b} \tau^{+} \tau^{-}$~\cite{CMS:2016knm} 
and $b \bar{b} \gamma \gamma$~\cite{CMS:2016vpz} to 300 fb$^{-1}$ of integrated luminosity, assuming 
a na\"ive $\sqrt{\mathcal{L}}$ (root-squared luminosity) improvement of the present CMS 95\% C.L. limit in both final states, 
and compare with the $b \bar{b} W^{+} W^{-}$ limits from Fig.~\ref{limits_mmcorno} (Right)\footnote[5]{We note 
the results from Fig.~\ref{limits_mmcorno} (Right) assume an eventual combination of CMS and ATLAS. This means that a $\sim \sqrt{2}$ sensitivity 
improvement should be added to the CMS $b \bar{b} \tau^{+} \tau^{-}$ and $b \bar{b} \gamma \gamma$ limits for a fair comparison. The comparison 
also assumes SM branching fractions for $h_1$, which is 
indeed the case for the xSM.}.
The results are shown in Fig.~\ref{limbbtautau}, and indicate that while $b \bar{b} \gamma \gamma$ provides the best limits for low $h_2$ masses
while $b \bar{b} W^{+} W^{-}$ may yield better limits than either $b \bar{b} \gamma \gamma$ or $b \bar{b} \tau^{+} \tau^{-}$ in the high mass region. 
We nevertheless stress that this comparison of $b \bar{b} \tau^{+} \tau^{-}$, $b \bar{b} \gamma \gamma$ and $b \bar{b} W^{+} W^{-}$ sensitivities is to be regarded 
as only indicative, 
since it is expected that future
sensitivity in the $b \bar{b} \tau^{+} \tau^{-}$ and $b \bar{b} \gamma \gamma$ final states improves better than $\sqrt{\mathcal{L}}$,
but a precise estimate of the comparative sensitivity is out of the scope of current work.
Still, the comparison suggests that $b \bar{b} W^{+} W^{-}$ is indeed a competitive search channel for resonant di-Higgs 
production at the LHC, particularly for high $m_2$ masses. 

\section{Outlook}
\label{sec:out}

Exploring the thermal history of EWSB is an important endeavor for particle physics and one for which high energy $pp$ collisions at the LHC and future colliders  
can provide invaluable input. Monte Carlo studies imply that for the mass of the observed Higgs boson, EWSB in the SM occurs through a 
cross over transition. However, the simplest extension of the SM scalar sector -- the xSM -- may lead to a decidedly different thermal history. 
In particular, for suitable choices of model parameters, the xSM can generate a strong first order EWPT, thereby fulfilling one of the key conditions 
for baryogenesis at the electroweak scale. Among the possible signatures of this possibility is resonant di-Higgs production in LHC $pp$ collisions, 
catalyzed by the interaction of the singlet-like scalar with pairs of the SM-like Higgs bosons.

In order to fully probe this possibility, it is important to consider a variety of possible final states associated with the di-Higgs decay 
products. Here, we have considered the $b{\bar b} W^+W^-$ channel, with the $W$-bosons decaying leptonically. The presence of two neutrinos 
in the final state makes the reconstruction of the decaying Higgs-like boson (and thus, of the parent singlet-like scalar) challenging. To address 
this challenge, we have developed a new Heavy Mass Estimator technique that allows one to achieve the needed mass reconstruction of the singlet-like scalar. 
Employing the HME and a MVA analysis of signal and background, we show that one is able to exclude the first order EWPT parameter space 
associated with maximum resonant di-Higgs production cross section with 300 fb$^{-1}$ of integrated luminosity for $m_2\lsim 700$ GeV and 
a statistically significant observation over roughly the same mass range with 3000 fb$^{-1}$. The projected sensitivity in the 
$b{\bar b} W^+W^-$ channel exceeds in the high mass region ($m_2\gsim 400$ GeV) that expected from $b{\bar b}\tau^+\tau^-$ and 
$b{\bar b}\gamma\gamma$ channels based on a na\"ive extrapolation of the present CMS 13 TeV public results for the latter channels, 
indicating that $b{\bar b} W^+W^-$ is a competitive di-Higgs LHC search channel for high invariant masses.

\vspace{2mm}

Finally, putting our results in the context of other, prospective future collider probes, we note that part, but not all, of the EWPT-viable xSM 
parameter space accessible in the $b{\bar b} W^+W^-$ channel at the LHC would be accessible with precision Higgs studies at the International 
Linear Collider with $\sqrt{s} = 1$ TeV and 1 ab$^{-1}$ integrated luminosity. Full access would require a circular $e^+e^-$ collider ($\sqrt{s}= 240$ GeV 
and 1 ab$^{-1}$)~\cite{Kotwal:2016tex}. Should the HL-LHC exclude this portion of parameter space, then a comprehensive probe would 
likely require a future 100 TeV $pp$ collider.


\acknowledgments

We thank Andreas Papaefstathiou for support with Herwig++. A.S. and T.H. are supported, in part, by the U.S. Department of Energy grant 
DE-SC0010813, A.S. and L.P. are supported in 
part by the Qatar National Research Foundation grant NPRP9-328-1-066.
M.J.R.M. and P.W. are supported, in part, under U.S. Department of Energy contract DE-SC0011095. 
J.M.N. is supported in part by the People Programme (Marie Curie Actions) of the European Union Seventh
Framework Programme (FP7/2007-2013), REA grant agreement
PIEF-GA-2013-625809 and the European Research Council under the
European Unions Horizon 2020 program (ERC Grant Agreement no.648680 DARKHORIZONS). 
J.M.N., M.J.R.M., M.S., and P.W. thank the Munich Institute for Astro- and Particle-Physics (MIAPP),
where a portion of this work has been completed. MS is supported in part by the European Commission through the "HiggsTools" Initial Training Network PITN-GA-2012-316704.
We finally thank the CMS collaboration for making {\sc Delphes} available.



\bibliographystyle{JHEP.bst}
\bibliography{bbWWRefs}  
  



%
\end{document}